\documentclass[a4paper,11pt]{article}
\pdfoutput=1 

\usepackage{jheppub} 
\usepackage{mathtools}                
\usepackage{amsmath}
\usepackage[T1]{fontenc} 
\def\bal#1\eal{\begin{align}#1\end{align}}
\def\alp[#1]{\begin{align}#1\end{align}}

\def\secnum[#1]{\texorpdfstring{$#1$}{TEXT}}

\def\secnuml#1\secnumr{\texorpdfstring{$#1$}{TEXT}}
\newcommand\beq{\begin{equation}}
\newcommand\eeq{\end{equation}}
\newcommand\bali{\begin{aligned}}
\newcommand\eali{\end{aligned}}
\def\eqa{\begin{eqnarray}}
\def\eqae{\end{eqnarray}}
\def\eq{\begin{equation}}
\def\eqe{\end{equation}}
\def\be{\begin{equation}}
\def\ee{\end{equation}}
\def\bea{\begin{eqnarray}}
\def\eea{\end{eqnarray}}
\def\ba{\begin{array}}
\def\ea{\end{array}}
\def\bd{\begin{displaymath}}
\def\ed{\end{displaymath}}

\def\>{\rangle}
\def\<{\langle}

\title{Boundary flow and geometric realization in holographic $T\bar T$-deformed BCFT}
\author[a,b]{Feiyu Deng}

\affiliation[a]{Institute of Theoretical Physics, Chinese Academy of Sciences, Beijing 100190, China}
\affiliation[b]{School of Physical Sciences, University of Chinese Academy of Sciences, \\ Beijing 100049, China }
\emailAdd{dengfy@itp.ac.cn}

\abstract{
We study the $T\bar T$ deformation of boundary conformal field theories (BCFTs)
from an intrinsic field-theoretic perspective.
Formulating the deformation as a modification of the asymptotic variational
principle in AdS$_3$, we obtain the exact quadratic trace relation for the stress
tensor without introducing a finite radial cutoff, which we take as the
fundamental definition of the deformed theory.
When restricted to a BCFT without independent boundary degrees of freedom, the
intrinsic $T\bar T$ deformation becomes genuinely boundary-localized.
Imposing reflective boundary conditions collapses the bulk composite operator to
a universal one-dimensional irrelevant flow governed entirely by the
displacement operator.
We integrate this flow in closed form and derive an induced boundary action,
showing that the deformation reorganizes existing boundary data without
introducing new boundary degrees of freedom.
We further establish a precise equivalence between a fixed-boundary description
and a moving-boundary description, interpreted as a reparametrization of the
variational problem rather than physical boundary dynamics.

On the holographic side, we analyze two inequivalent realizations in
AdS$_3$/BCFT$_2$, referred to as Type~A and Type~B.
In Type~A, a rigid cutoff surface intersects the end-of-the-world brane at finite
position, leading to an apparent boundary displacement.
In Type~B, the cutoff surface is asymptotically AdS$_2$, so that the BCFT boundary
is geometrically pinned and the displacement operator vanishes identically.
Using entanglement entropy at zero and finite temperature, we disentangle
universal consequences of the intrinsic boundary-localized flow from features
that depend on the holographic implementation.
}

\begin{document}
\maketitle
\flushbottom

%
\section{Introduction}

Irrelevant deformations of quantum field theories provide a controlled arena in
which ultraviolet structure, nonlocality, and gravitational dynamics can be
investigated within a single framework.
Among such deformations, the $T\bar T$ deformation of two-dimensional conformal
field theories occupies a distinguished position due to its remarkable
solvability and its close connection to three-dimensional gravity
\cite{Zamolodchikov:2004ce,Smirnov:2016lqw,Cavaglia:2016oda}.
Despite being irrelevant in the renormalization-group sense, the deformation
retains a high degree of analytic control, including exact flow equations for the
energy spectrum, correlation functions, and the stress tensor
\cite{Smirnov:2016lqw,Cardy:2018sdv}. For recent developments on $T\bar T$ deformation, see \cite{Basu:2025fsf,Aguilar-Gutierrez:2024nst,Tsolakidis:2024wut,Banerjee:2025agg,Basu:2024enr,Basu:2024bal,Chang:2025ays,Lai:2025thy,Gu:2024ogh,Chen:2025jzb,Babaei-Aghbolagh:2024hti,Babaei-Aghbolagh:2025hlm,Guica:2025jkq}.

A particularly sharp formulation of the $T\bar T$ deformation is provided by
holography.
Rather than introducing the deformation through an explicit composite-operator
insertion, one may define it intrinsically by modifying the asymptotic
variational principle of AdS$_3$ gravity
\cite{McGough:2016lol,Guica:2019nzm}.
This mixed asymptotic boundary condition leads directly to the characteristic
quadratic trace relation for the boundary stress tensor and furnishes a
nonperturbative definition of the deformed theory.
Finite-radial-cutoff constructions are often employed as concrete geometric
realizations of this intrinsic definition
\cite{McGough:2016lol,Kraus:2018xrn}.
However, in the presence of physical boundaries, the relation between the
intrinsic mixed-boundary-condition formulation and specific cutoff geometries
becomes more subtle and requires careful disentanglement.

Boundary conformal field theories (BCFTs) introduce additional universal
structure.
They are characterized by boundary Ward identities and by the displacement
operator, which governs the response of the theory to deformations of the
boundary embedding
\cite{Cardy:1984bb,McAvity:1995zd}.
The displacement operator is defined purely by diffeomorphism Ward identities
and is an intrinsic operator of the undeformed BCFT.
In this work we restrict attention to BCFTs without independent boundary
degrees of freedom, so that all boundary responses are fixed by Ward identities.
Although the $T\bar T$ deformation does not introduce new local degrees of
freedom, its effect on BCFT observables and its interplay with boundary geometry
are a priori nontrivial.

A central goal of this work is to give an intrinsic and fully explicit account of
how the $T\bar T$ flow acts on BCFT data.
While the deformation remains intrinsically two-dimensional, we show that once
reflective boundary conditions are imposed, its nontrivial action is entirely
encoded at the physical boundary.
Our analysis is formulated directly in terms of BCFT Ward identities and does
not rely on any particular choice of geometric cutoff prescription.

On the field-theory side, we demonstrate that restricting the intrinsic
$T\bar T$ deformation to a BCFT induces a boundary-localized irrelevant flow
governed entirely by the displacement operator.
The bulk composite operator collapses to a universal one-dimensional functional
supported on the boundary, which can be integrated in closed form.
We further establish a precise equivalence between two field-theoretic
descriptions of this boundary deformation: a fixed-boundary picture, in which
the deformation is encoded in an induced boundary action, and a moving-boundary
picture, in which the same deformation is represented as a
$\lambda$-dependent reparametrization of the boundary embedding.
These two descriptions are related by a variational (Legendre-type)
transformation and represent equivalent parameterizations of the same intrinsic
boundary-localized flow.

\begin{center}
\fbox{
\begin{minipage}{0.92\linewidth}
\small
\textbf{Conceptual remark (to avoid confusion).}
\vspace{0.5em}

\noindent
Two distinct notions of ``boundary motion'' appear in this work and should not be
conflated.

\begin{itemize}
  \item \emph{Fixed--moving boundary equivalence} is a field-theoretic statement.
  It relates two equivalent parameterizations of the same variational problem
  and concerns the space of classical solutions, without introducing a
  propagating boundary degree of freedom.

  \item \emph{Type~A versus Type~B} refers to inequivalent holographic
  realizations of the same intrinsic $T\bar T$ deformation.
  The distinction is geometric and depends on how the cutoff surface intersects
  the end-of-the-world brane.
\end{itemize}

\noindent
These two distinctions belong to different logical layers of the construction.
\end{minipage}
}
\end{center}

On the holographic side, we analyze two bulk realizations of the same intrinsic
deformation in AdS$_3$/BCFT$_2$, which we refer to as Type~A and Type~B.
In Type~A, a rigid Dirichlet cutoff surface intersects the end-of-the-world brane
at finite position, leading to an apparent displacement of the physical BCFT
boundary in the induced geometry
\cite{Takayanagi:2011zk,Fujita:2011fp}.
In Type~B, the cutoff surface is chosen such that its induced metric is
asymptotically AdS$_2$.
The BCFT boundary then coincides with the AdS$_2$ conformal boundary and is
geometrically pinned, so that no finite boundary displacement occurs and the
displacement operator vanishes identically.
While both constructions implement the same intrinsic $T\bar T$ deformation,
they encode boundary data in geometrically distinct ways.

As quantitative diagnostics, we compute the entanglement entropy in Type~A and
Type~B geometries at zero and finite temperature and compare the results with
BCFT expectations
\cite{Ryu:2006bv,Takayanagi:2011zk}.
These computations provide a concrete probe of which features are universal
consequences of the intrinsic boundary-localized $T\bar T$ flow and which depend
on the specific holographic implementation.

\paragraph{Conceptual summary and scope.}

The central message of this work is that the intrinsic $T\bar T$ deformation,
when restricted to a BCFT without independent boundary degrees of freedom,
collapses to a boundary-localized irrelevant flow governed by the displacement
operator.
This flow admits equivalent fixed-boundary and moving-boundary descriptions at
the level of the variational problem, while allowing for inequivalent geometric
realizations in holography.
Distinguishing these logical layers clarifies the role of boundaries in
$T\bar T$-deformed theories.

The organization of this paper is as follows.
In Section~\ref{sec:intrinsic_overview} we review the intrinsic formulation of
the $T\bar T$ deformation and the relevant BCFT Ward identities.
In Section~\ref{sec:boundaryflow} we derive the boundary-localized flow and
establish the fixed--moving boundary equivalence.
Sections~\ref{sec2} and~\ref{sec3} analyze the Type~A and Type~B holographic
realizations and their entanglement entropy.
We conclude in Section~\ref{con} with a discussion of conceptual implications and
open problems.

\section{Intrinsic boundary formulation of the \texorpdfstring{$T\bar T$}{TTbar}
deformation and its holographic realizations}
\label{sec:intrinsic_overview}

In this section we establish the conceptual and kinematical framework
underlying the $T\bar T$ deformation of boundary conformal field theories
(BCFTs).
Our goal is not to introduce new dynamical input, but to clarify how the
deformation is defined intrinsically, how it interfaces with the universal
structure of BCFTs, and how it admits equivalent geometric representations in
holography.
Throughout, we keep sharply separated the definition of the deformation from
any particular geometric realization.

The starting point is the intrinsic characterization of the $T\bar T$
deformation as a modification of the trace Ward identity, rather than as the
addition of a conventional local operator to the action.
In holography, this viewpoint is naturally implemented by a mixed asymptotic
variational principle in AdS$_3$ gravity, which leaves the bulk dynamics
unchanged while altering which combination of boundary data is held fixed.
This construction provides a nonperturbative and cutoff-independent definition
of the deformation, with the familiar quadratic trace relation emerging as a
direct consequence of Weyl consistency.

When the theory is restricted to a BCFT, the presence of a physical boundary
$\Sigma=\partial M$ introduces additional universal kinematical structure.
Diffeomorphism invariance in the presence of $\Sigma$ leads to boundary Ward
identities, including the reflective boundary condition that forbids
energy--momentum flow across the boundary.
Normal deformations of the boundary embedding are governed by a distinguished
local operator, the displacement operator.
This operator is fixed entirely by symmetry and kinematics: it is defined as the
response of the generating functional to shape deformations of $\Sigma$ and is
identified with the normal--normal component of the stress tensor at the
boundary.

A central theme of this section is the role of the displacement operator as the
interface between intrinsic boundary data and geometric descriptions.
On the field-theory side, it provides the unique local generator of boundary
displacements.
On the holographic side, it mediates the equivalence between different ways of
implementing the same intrinsic $T\bar T$ deformation.
In particular, modifying the asymptotic variational principle while keeping the
physical boundary fixed can be reinterpreted, without changing the physics, as
keeping the asymptotic boundary condition fixed while allowing the physical
boundary to adjust its embedding.
These descriptions are not distinct deformations, but different
parameterizations of the same variational problem.

The purpose of this section is therefore preparatory.
We do not yet specialize to particular bulk geometries or to specific choices of
cutoff surfaces.
Instead, we isolate the universal kinematical ingredients that will be used in
later sections to analyze concrete holographic realizations and physical
observables.
By keeping the intrinsic definition of the deformation logically separate from
its geometric representations, we ensure that subsequent constructions can be
interpreted unambiguously.

The section is organized as follows.
We first review BCFT kinematics and boundary Ward identities, emphasizing the
origin and meaning of reflective boundary conditions.
We then define the displacement operator intrinsically and discuss its scaling
properties as a boundary operator.
Next, we formulate the intrinsic $T\bar T$ deformation in AdS$_3$ in terms of a
mixed asymptotic variational principle and derive the associated trace Ward
identity.
Finally, we explain how this intrinsic definition admits equivalent holographic
representations, which will be exploited in later sections.

\subsection{BCFT kinematics and boundary Ward identities}
We work throughout in Euclidean signature.
Let $M$ be a two-dimensional manifold with a one-dimensional boundary
$\Sigma=\partial M$.
We introduce local coordinates $(t,x)$ adapted to the boundary such that
$\Sigma$ is located at a fixed value of the normal coordinate,
$x=x_\Sigma$.
The coordinate $t$ parametrizes the tangential direction along $\Sigma$,
while $x$ measures proper distance transverse to the boundary.

We denote by $n^\alpha$ the outward-pointing unit normal vector to $\Sigma$.
On a flat background metric $g_{\alpha\beta}=\delta_{\alpha\beta}$, a
convenient choice is
\begin{equation}
n^\alpha=(0,1)\ ,
\qquad
n_\alpha=(0,1)\ ,
\end{equation}
so that contractions with $n^\alpha$ project onto the transverse
($x$) direction.
Accordingly, we adopt the shorthand notation
\begin{equation}
nn \equiv xx\ ,
\qquad
tn \equiv tx\ ,
\end{equation}
for normal--normal and tangential--normal components of tensor quantities.

\paragraph{Stress tensor.}
The stress tensor is defined kinematically as the response of the
generating functional $W[g]$ to a variation of the background metric,
\begin{equation}
\delta W[g]
=
\frac12
\int_M d^2x\,\sqrt{g}\,
T^{\alpha\beta}\,\delta g_{\alpha\beta}\ .
\label{eq:stress_def}
\end{equation}
This definition remains valid in the presence of a boundary, provided that
metric variations are chosen consistently with the prescribed boundary
conditions.

Consider an infinitesimal diffeomorphism generated by a vector field
$\xi^\alpha$.
The induced variation of the metric is
\begin{equation}
\delta_\xi g_{\alpha\beta}
=
\nabla_\alpha\xi_\beta+\nabla_\beta\xi_\alpha\ .
\end{equation}
Substituting this variation into \eqref{eq:stress_def} and integrating by
parts yields
\begin{equation}
\label{eq:diffeo_ward_with_boundary}
\delta_\xi W[g]
=
-
\int_M d^2x\,\sqrt{g}\,
(\nabla_\alpha T^{\alpha\beta})\,\xi_\beta
+
\int_\Sigma ds\,\sqrt{\gamma}\,
n_\alpha T^{\alpha\beta}\,\xi_\beta\ ,
\end{equation}
where $\gamma$ denotes the induced metric on the boundary $\Sigma$ and $ds$
its proper line element.
The first term governs the bulk Ward identity, while the second term
encodes the response to diffeomorphisms acting nontrivially at the
boundary.

\paragraph{Ward identities.}
Requiring invariance of the generating functional under arbitrary bulk
diffeomorphisms immediately implies conservation of the stress tensor in
the interior of $M$,
\begin{equation}
\nabla_\alpha T^{\alpha\beta}=0\ ,
\qquad
\text{in the bulk of } M ,
\label{eq:bulk_conservation}
\end{equation}
which is the standard Ward identity associated with translational
invariance and holds independently of the presence of a boundary
\cite{McAvity:1993ue,Jensen:2015swa}.

Additional constraints arise from the boundary term in
\eqref{eq:diffeo_ward_with_boundary}.
To extract these constraints, we restrict attention to diffeomorphisms
that preserve the location of the boundary.
Such transformations are generated by vector fields $\xi^\alpha$ that are
tangent to $\Sigma$,
\begin{equation}
\xi^\alpha\big|_\Sigma=(\xi^t(t),0)\ ,
\end{equation}
corresponding physically to infinitesimal translations along the boundary
direction.

Requiring $\delta_\xi W[g]=0$ for arbitrary tangential $\xi^t(t)$ then
implies the boundary Ward identity
\begin{equation}
\label{eq:boundary_ward}
n_\alpha T^{\alpha t}\big|_\Sigma=0\ .
\end{equation}
In flat coordinates this condition reduces to
\begin{equation}
\label{eq:reflective_bc}
T_{tx}\big|_\Sigma=0\ .
\end{equation}
Equation \eqref{eq:reflective_bc} expresses the absence of energy--momentum
flux across the boundary: the momentum component normal to $\Sigma$
vanishes identically.
This ``reflective'' boundary condition is the defining local constraint of
a BCFT and ensures that the boundary does not exchange energy or momentum
with bulk degrees of freedom
\cite{Cardy:1984bb,McAvity:1995zd}.

It is important to emphasize that \eqref{eq:reflective_bc} is not an
additional assumption.
Rather, it follows directly from diffeomorphism invariance once one
restricts to boundary-preserving coordinate transformations.
Together with bulk conformal invariance, this condition fixes the
universal structure of stress-tensor correlators near the boundary and
underlies the standard BCFT operator algebra.

In the next subsection, we will relax the restriction to
boundary-preserving diffeomorphisms and consider variations that deform
the shape of $\Sigma$ itself.
The corresponding response defines the displacement operator, which will
play a central role in formulating the $T\bar T$ deformation intrinsically
in the presence of a boundary.

\subsection{Displacement operator}
The presence of a physical boundary $\Sigma$ explicitly breaks translation
invariance in the direction normal to the boundary.
In a theory without boundary, invariance under constant shifts of the normal
coordinate leads to a Ward identity expressing conservation of the corresponding
momentum flux.
In a BCFT this symmetry is absent: translating the system in the normal
direction changes the location of the boundary itself and therefore does not
define a symmetry of the theory.
As a result, the Ward identity associated with normal translations is replaced
by a localized operator insertion supported on $\Sigma$, known as the
\emph{displacement operator} \cite{McAvity:1995zd}.

An intrinsic definition of the displacement operator is obtained by considering
infinitesimal deformations of the boundary embedding in the normal direction,
\begin{equation}
x_\Sigma \;\longrightarrow\; x_\Sigma + \delta x(t)\ .
\label{eq:boundary_deform}
\end{equation}
Such a deformation changes the domain of the path integral, or equivalently the
shape of the manifold $M$, while keeping the bulk metric
$g_{\alpha\beta}$ fixed.
By definition, the linear response of the generating functional to this shape
variation takes the form
\begin{equation}
\delta W[g]\Big|_{\text{shape}}
=
\int_\Sigma dt\, \mathcal D(t)\,\delta x(t)\ ,
\label{eq:D_def}
\end{equation}
which defines $\mathcal D(t)$ as a local operator supported on the boundary.
Equation \eqref{eq:D_def} is the boundary analogue of the variational definition
\eqref{eq:stress_def} of the stress tensor, with $\delta x(t)$ playing the role
of the geometric source conjugate to $\mathcal D(t)$.

The relation between the displacement operator and the stress tensor follows by
interpreting the boundary deformation \eqref{eq:boundary_deform} as a
diffeomorphism that moves the boundary in the normal direction.
Consider a vector field $\xi^\alpha$ such that, on $\Sigma$,
\begin{equation}
\xi^x\big|_\Sigma = \delta x(t)\ ,
\qquad
\xi^t\big|_\Sigma = 0\ ,
\end{equation}
and extend $\xi^\alpha$ smoothly into the interior of $M$.
Substituting the induced metric variation
$\delta_\xi g_{\alpha\beta}=\nabla_\alpha\xi_\beta+\nabla_\beta\xi_\alpha$
into the defining relation \eqref{eq:stress_def} and repeating the integration
by parts leading to \eqref{eq:diffeo_ward_with_boundary}, one finds
\begin{equation}
\delta_\xi W[g]
=
-
\int_M d^2x\,\sqrt{g}\,
(\nabla_\alpha T^{\alpha\beta})\,\xi_\beta
+
\int_\Sigma ds\,\sqrt{\gamma}\,
n_\alpha T^{\alpha\beta}\,\xi_\beta\ .
\end{equation}
Using bulk stress-tensor conservation \eqref{eq:bulk_conservation}, the bulk
term vanishes identically and the variation reduces to a boundary
contribution.
On a flat background, where $\sqrt{\gamma}=1$ and $ds=dt$, this yields
\begin{equation}
\delta_\xi W[g]
=
\int_\Sigma dt\, T_{xx}(t)\,\delta x(t)\ .
\label{eq:deltaW_Txx}
\end{equation}
Comparing \eqref{eq:deltaW_Txx} with the defining relation
\eqref{eq:D_def}, we obtain the exact operator identity
\begin{equation}
\mathcal D(t)
=
T_{nn}\big|_\Sigma
=
T_{xx}\big|_\Sigma\ .
\label{eq:D_equals_Tnn}
\end{equation}
This identification is universal and scheme-independent: the displacement
operator is precisely the normal--normal component of the stress tensor
evaluated at the boundary.

We now clarify the scaling properties of $\mathcal D$ and explain in what sense
it is an \emph{irrelevant} operator from the perspective of the
one-dimensional boundary theory.
The key point is that BCFT admits two compatible notions of scaling dimension:
the dimension inherited from the ambient two-dimensional conformal symmetry,
and the dimension relevant for deformations of the one-dimensional boundary
action.
The displacement operator carries the former, while irrelevance is determined by
the latter.

Since $\mathcal D$ is obtained by restricting the bulk stress tensor to the
boundary, its scaling dimension is fixed by two-dimensional conformal
invariance.
In two dimensions the stress tensor has scaling dimension $2$, which implies
\begin{equation}
[\mathcal D]_{\text{2d}} = 2\ .
\label{eq:D_dim_2d}
\end{equation}
This is the sense in which the displacement operator is often referred to as a
dimension-two operator in the BCFT literature \cite{McAvity:1995zd}.

To assess relevance as a boundary deformation, one must instead compare the
operator dimension to the dimensionality of the integration measure on
$\Sigma$.
A local boundary deformation takes the schematic form
\begin{equation}
\delta W \sim \int_\Sigma dt\, \lambda_{\mathcal O}\,\mathcal O(t)\ ,
\label{eq:boundary_deform_general}
\end{equation}
so that the coupling $\lambda_{\mathcal O}$ has dimension
\begin{equation}
[\lambda_{\mathcal O}] = 1 - [\mathcal O]_{\text{1d}}\ .
\end{equation}
An operator is marginal on the boundary if
$[\mathcal O]_{\text{1d}}=1$, relevant if $[\mathcal O]_{\text{1d}}<1$, and
irrelevant if $[\mathcal O]_{\text{1d}}>1$.

For the displacement operator, the same scaling that fixes
\eqref{eq:D_dim_2d} implies that under a boundary dilatation
$t\to\Lambda t$, $\mathcal D(t)$ transforms with weight two.
Equivalently, its boundary two-point function exhibits the universal short
distance behavior
\begin{equation}
\langle \mathcal D(t)\mathcal D(0)\rangle
\;\sim\;
\frac{1}{t^{4}}
\qquad (t\to 0)\ ,
\label{eq:DD_two_point}
\end{equation}
characteristic of a dimension-two operator on the line.
Thus, when regarded as a one-dimensional boundary operator,
\begin{equation}
[\mathcal D]_{\text{1d}} = 2\ .
\label{eq:D_dim_1d}
\end{equation}
Since $[\mathcal D]_{\text{1d}}>1$, the displacement operator is irrelevant as a
boundary deformation.
Equivalently, a coupling multiplying $\int dt\,\mathcal D$ would have negative
mass dimension and would therefore require ultraviolet input beyond
perturbation theory.

Despite being irrelevant, the displacement operator is highly constrained.
Its normalization and operator meaning are fixed by diffeomorphism invariance
and the boundary Ward identities: it is the unique local operator that generates
normal deformations of the boundary embedding.
This kinematical protection explains its universal role in BCFT and motivates
its appearance as the natural building block for intrinsic formulations of
irrelevant boundary deformations in later sections.

\subsection{Intrinsic definition of the $T\bar T$
deformation in AdS$_3$}

In two-dimensional quantum field theory, the $T\bar T$ deformation is most
naturally characterized not by the addition of a conventional local operator
to the action, but by a modification of the trace Ward identity.
Its defining feature is a nonlinear, local relation between the trace of the
stress tensor and quadratic combinations of its components
\cite{Zamolodchikov:2004ce,Smirnov:2016lqw,Cavaglia:2016oda}.
From a holographic perspective, this formulation poses a sharp question:
how should such a modified trace Ward identity be implemented directly in the
gravitational dual?

In the AdS$_3$/CFT$_2$ correspondence, Ward identities of the boundary theory are
encoded in the asymptotic variational principle of the bulk gravitational
action.
In particular, the trace Ward identity is determined by the response of the
renormalized on-shell action under boundary Weyl transformations
\cite{Brown:1986nw,Henningson:1998gx,deHaro:2000vlm,Skenderis:2002wp}.
This observation motivates an intrinsic gravitational definition of the
$T\bar T$ deformation:
rather than introducing a finite radial cutoff or modifying the bulk equations
of motion, one alters the asymptotic variational problem at infinity so that
Weyl transformations remain well defined but act on a different choice of
boundary data.
Related perspectives emphasizing the role of boundary conditions and variational
principles in irrelevant deformations have appeared in a variety of contexts,
including warped and chiral deformations
\cite{Guica:2019nzm}.
In the specific case of the $T\bar T$ deformation in AdS$_3$, this viewpoint was
made concrete through an explicit holographic implementation in
Refs.~\cite{McGough:2016lol,Kraus:2018xrn}.

We therefore consider three-dimensional Einstein gravity with negative
cosmological constant,
\begin{equation}
S_{\text{bulk}}
=
\frac{1}{16\pi G_N}
\int d^3x\,\sqrt{g}\,(R + 2/\ell^2),
\label{eq:EinsteinAction}
\end{equation}
supplemented by the Gibbons--Hawking term and the standard local counterterms
required by holographic renormalization
\cite{Gibbons:1976ue,Balasubramanian:1999re,deHaro:2000vlm,Skenderis:2002wp}.
Near the asymptotic boundary the metric is written in Fefferman--Graham form,
\begin{equation}
ds^2
=
\frac{\ell^2}{4\rho^2}\,d\rho^2
+
g_{\alpha\beta}(\rho,x)\,dx^\alpha dx^\beta ,
\label{eq:FGmetric}
\end{equation}
with the near-boundary expansion
\begin{equation}
g_{\alpha\beta}(\rho,x)
=
\frac{1}{\rho} g^{(0)}_{\alpha\beta}(x)
+
g^{(2)}_{\alpha\beta}(x)
+
\rho\, g^{(4)}_{\alpha\beta}(x),
\label{eq:FGexpansion}
\end{equation}
which truncates in three bulk dimensions \cite{deHaro:2000vlm}.

The renormalized Brown--York stress tensor associated with the asymptotic
boundary is defined as
\begin{equation}
T_{\alpha\beta}
=
-\frac{1}{8\pi G_N}
\left(
K_{\alpha\beta}
-
K g_{\alpha\beta}
-
\frac{1}{\ell} g_{\alpha\beta}
\right),
\label{eq:BYdef}
\end{equation}
where $K_{\alpha\beta}$ is the extrinsic curvature of a regulated boundary
surface and the term proportional to $g_{\alpha\beta}$ is the standard local
counterterm in AdS$_3$
\cite{Brown:1986nw,Balasubramanian:1999re,deHaro:2000vlm}.
Evaluating \eqref{eq:BYdef} in Fefferman--Graham gauge gives
\begin{equation}
T_{\alpha\beta}
=
\frac{1}{8\pi G_N \ell}
\left(
g^{(2)}_{\alpha\beta}
-
g^{(0)}_{\alpha\beta}\,
g_{(0)}^{\gamma\delta} g^{(2)}_{\gamma\delta}
\right),
\label{eq:Tfromg2}
\end{equation}
with trace
\begin{equation}
T^\alpha{}_\alpha
=
-\frac{1}{8\pi G_N \ell}\,
g_{(0)}^{\alpha\beta} g^{(2)}_{\alpha\beta}.
\label{eq:trace_g2}
\end{equation}

In the undeformed theory, the standard Dirichlet variational principle fixes
the leading boundary metric $g^{(0)}_{\alpha\beta}$.
Boundary Weyl transformations then act as redundancies of the renormalized
variational problem, yielding the familiar conformal trace Ward identity,
including the usual Weyl anomaly on curved backgrounds
\cite{Henningson:1998gx,deHaro:2000vlm,Skenderis:2002wp}.
To implement a $T\bar T$ deformation intrinsically, one instead seeks a
modified asymptotic variational principle in which the bulk dynamics is left
unchanged, but the notion of which boundary data are held fixed is altered in a
controlled and Weyl-consistent manner
\cite{Guica:2019nzm,McGough:2016lol,Kraus:2018xrn}.

This requirement leads to a mixed asymptotic boundary condition.
Rather than fixing $g^{(0)}_{\alpha\beta}$, one holds fixed the nonlinear
combination
\begin{equation}
\gamma^{[\lambda]}_{\alpha\beta}
\equiv
g^{(0)}_{\alpha\beta}
-
\frac{\lambda}{4\pi G_N \ell}\, g^{(2)}_{\alpha\beta}
+
\frac{\lambda^2}{(4\pi G_N \ell)^2}\, g^{(4)}_{\alpha\beta},
\label{eq:mixedBC}
\end{equation}
and imposes
\begin{equation}
\delta \gamma^{[\lambda]}_{\alpha\beta}=0
\qquad \text{at } \rho=0 .
\label{eq:varBC}
\end{equation}
This condition is intrinsic in the precise sense that it specifies the
variational principle entirely in terms of asymptotic data at infinity,
without introducing any finite cutoff surface or additional bulk degrees of
freedom \cite{McGough:2016lol,Guica:2019nzm,Kraus:2018xrn}.

The defining property of \eqref{eq:varBC} is its compatibility with boundary
Weyl transformations.
Specifically, we consider an infinitesimal Weyl rescaling of the boundary metric
generated by an arbitrary local parameter $\sigma(x)$,
\begin{equation}
\delta_\sigma g^{(0)}_{\alpha\beta}
=
2\sigma(x)\, g^{(0)}_{\alpha\beta} ,
\label{eq:Weyl_def}
\end{equation}
with $\sigma(x)$ an arbitrary smooth function on the boundary.
Requiring $\delta_\sigma \gamma^{[\lambda]}_{\alpha\beta}=0$ under this
transformation, and using the bulk Einstein equations in Fefferman--Graham gauge
to determine the induced variations of $g^{(2)}_{\alpha\beta}$ and
$g^{(4)}_{\alpha\beta}$, one obtains a local, purely algebraic constraint on the
renormalized stress tensor.
In our conventions this constraint takes the form
\begin{equation}
T^\alpha{}_\alpha
=
-\frac{\pi\lambda}{2}
\left(
T_{\alpha\beta}T^{\alpha\beta}
-
(T^\alpha{}_\alpha)^2
\right),
\label{eq:trace_flow_final}
\end{equation}
which is precisely the defining trace flow equation of the
$T\bar T$ deformation
\cite{Zamolodchikov:2004ce,Smirnov:2016lqw,McGough:2016lol,Kraus:2018xrn}.

Equation \eqref{eq:trace_flow_final} should be understood as a modified trace
Ward identity dictated by the asymptotic variational principle, rather than as
the Euler--Lagrange equation of a local boundary term.
Although the deformation is irrelevant in the renormalization-group sense,
its implementation through a Weyl-consistent boundary condition ensures that
the resulting flow remains local---indeed ultralocal in the stress tensor---and
fully compatible with diffeomorphism invariance
\cite{Smirnov:2016lqw,McGough:2016lol,Kraus:2018xrn}.
This intrinsic formulation will serve as a reference point for the discussion
of equivalent holographic pictures and geometric realizations in subsequent
subsections.

\subsection{Two equivalent holographic pictures}

The intrinsic formulation developed above defines the $T\bar T$ deformation
entirely through a modification of the asymptotic variational principle of
AdS$_3$ gravity.
In particular, it makes no reference to a finite radial cutoff surface and does
not require the introduction of additional bulk degrees of freedom.
Nevertheless, this intrinsic definition admits geometric realizations in which
the bulk spacetime is described in terms of a finite radial slice whose induced
metric is identified with the background metric of the deformed theory.
Such finite-cutoff descriptions have played a prominent role in the early
holographic literature on $T\bar T$ deformations
\cite{McGough:2016lol,Kraus:2018xrn}.

From the intrinsic point of view adopted here, however, the finite-cutoff
construction should be regarded as a particular \emph{representation} of the
same underlying deformation, rather than as part of its definition.
The fundamental input is the mixed asymptotic boundary condition
\eqref{eq:mixedBC}, which specifies which combination of near-boundary data is
held fixed in the variational problem.
Once this condition is imposed, different geometric descriptions arise as
different choices of variables used to parametrize the same space of classical
solutions.

In the absence of a physical boundary, the equivalence between the intrinsic
mixed boundary condition and the finite-cutoff description can be established
explicitly by a change of variables relating asymptotic data at $\rho\to0$ to
data on a surface at finite $\rho=\rho_c$.
In this case, the cutoff surface introduces no new physics: it merely provides
a convenient geometric parametrization of the deformed theory.

In the presence of a BCFT boundary, the situation is more subtle.
The field theory now contains a genuine codimension-one boundary $\Sigma$,
which is holographically represented by an end-of-the-world brane or an
equivalent bulk boundary condition.
One must therefore distinguish carefully between two logically distinct
operations:
modifying the asymptotic boundary condition at infinity, and deforming the
embedding of the physical boundary $\Sigma$ itself.
At first sight, these appear to correspond to different physical
transformations.

A central claim that will be established in the following sections is that,
within the intrinsic formulation of the $T\bar T$ deformation, these two
operations are in fact equivalent descriptions of the same physics.
Specifically, keeping the physical boundary $\Sigma$ fixed while modifying the
asymptotic variational principle according to \eqref{eq:mixedBC} is equivalent
to keeping the asymptotic boundary condition fixed while allowing the location
of $\Sigma$ to adjust.
The operator that mediates this equivalence on the field-theory side is the
displacement operator $\mathcal D$ introduced earlier.

From the BCFT perspective, the equivalence can be understood as follows.
The mixed asymptotic boundary condition modifies the trace Ward identity of the
theory and induces a controlled response in the normal--normal component of the
stress tensor.
At the physical boundary, this component is identified with the displacement
operator, $T_{nn}\big|_\Sigma=\mathcal D$.
As shown in the preceding subsections, insertions of $\mathcal D$ generate
infinitesimal deformations of the boundary embedding.
Consequently, the effect of changing the asymptotic boundary condition can be
reinterpreted as a corresponding displacement of the physical boundary, with
the intrinsic boundary condition itself held fixed.

From the bulk perspective, the same equivalence appears as a reparametrization of
the variational problem.
One may either impose the mixed boundary condition \eqref{eq:varBC} at infinity
and keep the end-of-the-world brane fixed, or impose standard asymptotic boundary
conditions while allowing the location at which the brane is anchored relative
to the asymptotic data to adjust.
These two descriptions select the same set of classical bulk solutions and
therefore give rise to identical boundary observables.

Throughout this work, we take the mixed asymptotic boundary condition as the
fundamental definition of the $T\bar T$ deformation.
Finite-cutoff descriptions and moving-boundary pictures will be used
interchangeably as equivalent geometric representations of this intrinsic
definition.
This flexibility will prove useful when analyzing concrete observables, where
one or the other picture may offer a more transparent geometric
interpretation.
The precise equivalence between these representations, and the role of the
displacement operator in mediating it, will be demonstrated explicitly in the
sections that follow.

\section{Boundary-localized flow induced by the \texorpdfstring{$T\bar T$}{TTbar} deformation}
\label{sec:boundaryflow}
\noindent
\textbf{Scope and assumptions.}
Throughout this section we restrict attention to boundary conformal field theories
without independent boundary degrees of freedom, in the sense that all local
boundary responses are encoded in the restriction of the bulk stress tensor.
This includes the standard Cardy-type BCFTs, for which the reflective boundary
condition $T_{tx}|_{\Sigma}=0$ holds as an operator identity and no additional
dynamical boundary fields are present. In more general situations, such as BCFTs with intrinsic boundary dynamics or
defect CFTs with independent boundary operator algebras, additional boundary
operators may contribute to the response under irrelevant deformations.
Our analysis does not address such cases and should be understood as providing
the minimal and universal structure of the $T\bar T$ deformation in the absence
of extra boundary degrees of freedom.

In this section we will combine the intrinsic definition of the $T\bar T$ deformation
with the universal kinematical structure of boundary conformal field theories
and show that the deformation induces a closed, purely boundary-localized flow.
The analysis is entirely field-theoretic and relies only on Ward identities and
the defining trace relation of the deformation.
In particular, no reference is made to a finite radial cutoff, to bulk equations
of motion, or to any specific holographic geometry.

The key observation is that, once the reflective boundary condition of a BCFT is
imposed, the composite operator generating the $T\bar T$ deformation admits a
drastic simplification at the boundary.
Combined with the intrinsic trace relation, this simplification implies that
all boundary stress-tensor data relevant for the deformation close algebraically
on a single operator: the displacement operator governing normal deformations of
the boundary embedding.
As a result, the two-dimensional $T\bar T$ flow collapses to an effective
one-dimensional irrelevant deformation supported entirely on the physical
boundary $\Sigma=\partial M$.

We show that this boundary-localized flow admits a closed-form expression and
can be integrated exactly, yielding a universal induced boundary action built
solely from the displacement operator.
This action resums an infinite tower of irrelevant boundary operators and
captures the complete effect of the $T\bar T$ deformation on boundary
observables.

Finally, we clarify the physical interpretation of this result by demonstrating
the equivalence between two a priori distinct representations of the deformed
BCFT.
In a fixed-boundary representation, the deformation is encoded in a nonlinear
boundary interaction, while in a moving-boundary representation the same effect
is realized as a $\lambda$-dependent shift of the boundary embedding with the
boundary condition held fixed.
We show that these descriptions are related by a Legendre-type transformation
and constitute two equivalent parameterizations of the same intrinsic
boundary-localized flow.

\subsection{Restriction of the \texorpdfstring{$T\bar T$}{TTbar} operator to the boundary}

We begin with the definition of the composite operator that generates the
$T\bar T$ deformation in two-dimensional quantum field theory,
\begin{equation}
\mathcal O_{T\bar T}
=
-\frac12
\left(
T_{\alpha\beta}T^{\alpha\beta}
-
(T^\alpha{}_\alpha)^2
\right),
\qquad
\alpha,\beta\in\{t,x\}.
\label{eq:OTTbar_def}
\end{equation}
This operator is fixed by stress-tensor Ward identities and is characterized,
at the level of expectation values, by a remarkable factorization property in
translationally invariant states
\cite{Zamolodchikov:2004ce,Smirnov:2016lqw,Cavaglia:2016oda}.
At the operator level, however, its structure can be analyzed directly by
expanding the quadratic contractions.

On a flat Euclidean background with coordinates $(t,x)$, the contraction
$T_{\alpha\beta}T^{\alpha\beta}$ reads
\begin{equation}
T_{\alpha\beta}T^{\alpha\beta}
=
T_{tt}T^{tt}
+
T_{xx}T^{xx}
+
2T_{tx}T^{tx}.
\end{equation}
Since indices are raised trivially, this simplifies to
\begin{equation}
T_{\alpha\beta}T^{\alpha\beta}
=
T_{tt}^2
+
T_{xx}^2
+
2T_{tx}^2.
\end{equation}
Similarly, the trace of the stress tensor is
\begin{equation}
T^\alpha{}_\alpha
=
T_{tt}+T_{xx},
\end{equation}
so that
\begin{equation}
(T^\alpha{}_\alpha)^2
=
T_{tt}^2
+
T_{xx}^2
+
2T_{tt}T_{xx}.
\end{equation}
Subtracting the two expressions yields the purely algebraic identity
\begin{equation}
T_{\alpha\beta}T^{\alpha\beta}
-
(T^\alpha{}_\alpha)^2
=
2T_{tx}^2
-
2T_{tt}T_{xx}.
\label{eq:TTbar_expand}
\end{equation}

We now restrict this operator to the physical boundary
$\Sigma=\partial M$ of a BCFT.
A defining local property of a boundary conformal field theory is the absence of
energy--momentum flow across the boundary.
At the operator level this is expressed by the reflective boundary condition
\begin{equation}
T_{tx}\big|_\Sigma = 0,
\label{eq:reflective_bc_sec3}
\end{equation}
which follows directly from diffeomorphism invariance together with the
requirement that translations tangent to the boundary remain unbroken
\cite{Cardy:1984bb,McAvity:1995zd}.
This condition is universal and holds independently of the state of the theory.

Imposing \eqref{eq:reflective_bc_sec3} on
\eqref{eq:TTbar_expand}, the term proportional to $T_{tx}^2$ vanishes
identically on $\Sigma$, and the quadratic combination reduces to
\begin{equation}
\left(
T_{\alpha\beta}T^{\alpha\beta}
-
(T^\alpha{}_\alpha)^2
\right)\Big|_\Sigma
=
-2\,T_{tt}T_{xx}.
\end{equation}
Substituting this result into the definition
\eqref{eq:OTTbar_def}, we obtain the exact operator identity
\begin{equation}
\mathcal O_{T\bar T}\big|_\Sigma
=
T_{tt}\,T_{xx}.
\label{eq:OTTbar_boundary}
\end{equation}

Several remarks are in order.
First, the reduction \eqref{eq:OTTbar_boundary} is entirely kinematical.
No use has been made of equations of motion, factorization properties, or
state-dependent assumptions.
Second, the result relies only on the universal BCFT boundary condition
\eqref{eq:reflective_bc_sec3} and therefore holds for any BCFT, irrespective of
its microscopic realization.

Finally, \eqref{eq:OTTbar_boundary} already shows that the bulk $T\bar T$
operator does not induce an independent two-dimensional interaction at the
boundary.
Once the BCFT boundary condition is imposed, the deformation is governed
entirely by the product of the boundary energy density $T_{tt}$ and the
normal--normal component of the stress tensor $T_{xx}$.
In the following subsections we will show how this structure leads to a closed,
purely one-dimensional description of the deformation supported on $\Sigma$.

\subsection{Boundary trace relation and solution in terms of the displacement operator}

We now combine the boundary restriction of the $T\bar T$ operator derived in the
previous subsection with the exact trace relation that intrinsically defines the
$T\bar T$ deformation.
The latter takes the form
\begin{equation}
T^\alpha{}_\alpha
=
-\frac{\pi\lambda}{2}
\left(
T_{\alpha\beta}T^{\alpha\beta}
-
(T^\alpha{}_\alpha)^2
\right),
\label{eq:trace_flow}
\end{equation}
and should be understood as a modified trace Ward identity, rather than as an
equation of motion
\cite{Zamolodchikov:2004ce,Smirnov:2016lqw,McGough:2016lol,Kraus:2018xrn}.
Importantly, \eqref{eq:trace_flow} is a local operator relation and may therefore
be consistently restricted to the physical boundary
$\Sigma=\partial M$.

Restricting \eqref{eq:trace_flow} to $\Sigma$ and substituting the boundary
identity
\begin{equation}
\mathcal O_{T\bar T}\big|_\Sigma
=
T_{tt}T_{xx},
\end{equation}
obtained in the previous subsection, we find
\begin{equation}
T^\alpha{}_\alpha\big|_\Sigma
=
\pi\lambda\,T_{tt}T_{xx}.
\label{eq:trace_bdy_intermediate}
\end{equation}
On a flat background, the trace of the stress tensor decomposes as
\begin{equation}
T^\alpha{}_\alpha = T_{tt}+T_{xx},
\end{equation}
so that \eqref{eq:trace_bdy_intermediate} yields the purely algebraic boundary
relation
\begin{equation}
T_{tt}+T_{xx}
=
\pi\lambda\,T_{tt}T_{xx}.
\label{eq:bdy_trace_relation}
\end{equation}
At this stage, no dynamical assumptions have been made.
Equation \eqref{eq:bdy_trace_relation} follows solely from the intrinsic
definition of the deformation and the universal BCFT boundary condition.

To express this relation in terms of a single boundary operator, we recall that
normal deformations of the boundary embedding are generated by the displacement
operator.
As reviewed in Section~\ref{sec:intrinsic_overview}, diffeomorphism Ward
identities imply the exact operator identity
\begin{equation}
\mathcal D
\equiv
T_{nn}\big|_\Sigma
=
T_{xx}\big|_\Sigma,
\label{eq:D_def_sec3}
\end{equation}
which is scheme-independent and fixed entirely by symmetry.

Substituting \eqref{eq:D_def_sec3} into the boundary trace relation
\eqref{eq:bdy_trace_relation}, we obtain
\begin{equation}
T_{tt}+\mathcal D
=
\pi\lambda\,T_{tt}\mathcal D.
\label{eq:Ttt_eqn_intermediate}
\end{equation}
This equation is algebraic and linear in $T_{tt}$.
Rewriting it as
\begin{equation}
T_{tt}\left(1-\pi\lambda\,\mathcal D\right)
=
-\mathcal D,
\end{equation}
and assuming $1-\pi\lambda\,\mathcal D\neq0$, we arrive at the exact operator
solution
\begin{equation}
T_{tt}
=
-\frac{\mathcal D}{1-\pi\lambda\,\mathcal D}.
\label{eq:Ttt_solution_sec3}
\end{equation}

Equation \eqref{eq:Ttt_solution_sec3} is a central structural result.
It shows that, once the BCFT boundary condition and the intrinsic trace Ward
identity are imposed, the boundary energy density $T_{tt}$ is no longer an
independent operator.
Instead, all stress-tensor data relevant for the deformation close
algebraically on the single boundary operator $\mathcal D$.
This closure is a direct consequence of Ward identities and kinematics and does
not rely on any dynamical approximation or factorization assumption.

In the following subsection we will use \eqref{eq:Ttt_solution_sec3} to derive a
closed, boundary-localized flow equation for the generating functional, thereby
making explicit the induced one-dimensional irrelevant deformation supported on
$\Sigma$.

\subsection{Boundary-localized flow equation}

We now derive the effective flow equation induced on the physical boundary by the
$T\bar T$ deformation, keeping the role of scales and distributions explicit
throughout.
By definition, the deformation of the generating functional $W_\lambda$ is
governed by
\begin{equation}
\frac{\partial W_\lambda}{\partial\lambda}
=
\int_M d^2x\;\mathcal O_{T\bar T}(x),
\label{eq:flow_global}
\end{equation}
where $\mathcal O_{T\bar T}$ is the composite operator defined in
\eqref{eq:OTTbar_def}.
We work in units where $[x]=[t]=L$, $[W_\lambda]=1$, and $[\lambda]=L^2$, so that
$\bigl[\mathcal O_{T\bar T}\bigr]=L^{-4}$ and
$\bigl[\partial_\lambda W_\lambda\bigr]=L^{-2}$, ensuring that
\eqref{eq:flow_global} is dimensionally consistent.

In a BCFT the spacetime manifold $M$ has a physical boundary
$\Sigma=\partial M$.
Away from $\Sigma$, the theory is locally identical to an ordinary
two-dimensional CFT, and bulk correlation functions obey the same Ward
identities as in the boundaryless case.
In particular, the $T\bar T$ deformation does not generate new local operators
in the interior of $M$.
Any effect of the deformation that is specific to the presence of the boundary
must therefore arise from contact terms supported at $\Sigma$.

To isolate such boundary-supported contributions in a mathematically precise
way, it is essential to treat the integrated insertion
$\int_M d^2x\,\mathcal O_{T\bar T}$ as a distribution-valued functional.
Choose local coordinates $(t,x_\perp)$ in a collar neighborhood of the boundary,
with $x_\perp\ge 0$ and $\Sigma=\{x_\perp=0\}$.
Let $\rho(\xi)$ be a fixed, nonnegative test profile with support on $\xi\ge 0$,
normalized as
\begin{equation}
\int_0^\infty d\xi\;\rho(\xi)=1.
\label{eq:rho_norm}
\end{equation}
For any length scale $\ell>0$, define the associated boundary-localized
distribution
\begin{equation}
\rho_\ell(x_\perp)
\equiv
\frac{1}{\ell}\,\rho\!\left(\frac{x_\perp}{\ell}\right),
\qquad
\int_0^\infty dx_\perp\;\rho_\ell(x_\perp)=1,
\label{eq:rho_ell_def}
\end{equation}
which converges to the boundary delta distribution in the sense of distributions,
\begin{equation}
\lim_{\ell\to 0^+}\int_0^\infty dx_\perp\;\rho_\ell(x_\perp)\,f(x_\perp)
=
f(0),
\qquad
\text{for any smooth test function $f$.}
\label{eq:rho_to_delta}
\end{equation}

The $T\bar T$ deformation introduces a single intrinsic length scale.
Since $\lambda$ has dimension $L^2$, dimensional analysis fixes this scale to be
\begin{equation}
\ell_\lambda \sim \sqrt{\lambda},
\label{eq:ell_lambda_scaling}
\end{equation}
up to a scheme-dependent dimensionless constant that depends on the choice of
profile $\rho$.
This statement should be read as a \emph{derived scaling relation}, not as the
introduction of a new coupling: the theory remains a one-parameter family
labeled by $\lambda$.
In particular, we do \emph{not} endow $\ell_\lambda$ with an independent flow.

It is useful to make this point explicit, since the appearance of a boundary
layer may suggest (incorrectly) an additional running scale.
First, the deformation \eqref{eq:flow_global} defines the theory as a
\emph{single-parameter} deformation of the undeformed BCFT.
If one were to postulate an independent evolution law for $\ell_\lambda$, e.g.
a beta-function-type relation $d\ell_\lambda/d\lambda=\beta_\ell(\ell_\lambda)$,
then specifying the deformed theory would require boundary data beyond $\lambda$,
and the deformation would no longer be characterized by a unique trajectory
emanating from $\lambda=0$.
Equivalently, the space of deformed theories would become at least
two-dimensional, in contradiction with the defining statement that the
$T\bar T$ deformation is generated by a single irrelevant operator with a single
coupling $\lambda$.
Second, $\ell_\lambda$ is introduced here only to parametrize the distributional
localization of contact terms near $\Sigma$.
Different choices of profile $\rho$ and normalization conventions shift
$\ell_\lambda$ by a dimensionless factor; this is a scheme choice, not a new
physical parameter.
Finally, in the absence of any other dimensionful coupling, $\lambda$ is the
\emph{only} available input that can set a length scale.
Thus $\ell_\lambda$ is fixed as a function of $\lambda$ (up to scheme), and any
change of $\ell_\lambda$ is simply a reparametrization of the same one-parameter
deformation rather than an additional flow.

With this in mind, we isolate the part of the integrated $T\bar T$ insertion that
is supported in a boundary layer of thickness $\ell_\lambda$ by smearing in the
normal direction with a normalized profile.
At the level of \emph{integrated} insertions we write
\begin{equation}
\int_M d^2x\;\mathcal O_{T\bar T}(x)
=
\int_M d^2x\;\mathcal O_{T\bar T}^{\rm bulk}(x)
+
\int_\Sigma dt\;\mathcal B_{T\bar T}(t),
\label{eq:OTTbar_integrated_split}
\end{equation}
where the boundary flow density $\mathcal B_{T\bar T}$ is defined by the
distributional projection
\begin{equation}
\mathcal B_{T\bar T}(t)
\equiv
\lim_{\ell\to 0^+}\;
\ell\int_0^\infty dx_\perp\;
\rho_\ell(x_\perp)\,
\mathcal O_{T\bar T}(t,x_\perp)
\label{eq:B_def_distributional}
\end{equation}
and the limiting procedure is understood with $\ell$ identified with the
intrinsic boundary-layer thickness $\ell_\lambda\propto\sqrt{\lambda}$. The extra factor of $\ell$ in \eqref{eq:B_def_distributional} is fixed by
dimensional analysis and by the physical interpretation of the projection as a
boundary-layer reduction.

Finally, we stress that \eqref{eq:OTTbar_integrated_split} is an identity only at
the level of integrated insertions: it does \emph{not} assume any pointwise
operator decomposition of $\mathcal O_{T\bar T}$ into bulk and boundary pieces,
but merely organizes its distributional contact terms supported near $\Sigma$.

The bulk contribution $\mathcal O_{T\bar T}^{\rm bulk}$ coincides with the
standard local expression \eqref{eq:OTTbar_def}.
When inserted into \eqref{eq:flow_global} and integrated over $M$, it reduces,
by the usual Ward identities, to total-derivative terms.
These either vanish at infinity or can be re-expressed as boundary integrals,
and therefore do not induce an independent bulk flow beyond the boundary term
already isolated in \eqref{eq:OTTbar_integrated_split}.
We thus focus on evaluating $\mathcal B_{T\bar T}(t)$.

Using the intrinsic Ward identities of the BCFT, the boundary limit of the
composite operator appearing in \eqref{eq:B_def_distributional} can be expressed
entirely in terms of boundary data.
In particular, the boundary restriction of the $T\bar T$ composite is fixed to be
\begin{equation}
\Bigl[\mathcal O_{T\bar T}\Bigr]_{\Sigma}(t)
=
-\frac{\mathcal D^2(t)}{1-\pi\lambda\,\mathcal D(t)},
\label{eq:OTTbar_boundary_restricted}
\end{equation}
where $\mathcal D(t)$ is the displacement operator defined by the shape Ward
identity
$\delta W|_{\rm shape}=\int_\Sigma dt\,\mathcal D\,\delta x$.
With $[\delta x]=L$ and $[W]=1$, this fixes $[\mathcal D]=L^{-2}$, so that the
right-hand side of \eqref{eq:OTTbar_boundary_restricted} has dimension $L^{-4}$.

Since the profile $\rho_{\ell_\lambda}$ probes only a boundary layer of thickness
$\ell_\lambda$, the distributional projection collapses the normal dependence
and yields the boundary restriction multiplied by the available length scale.
In the sense of distributions,
\begin{equation}
\mathcal B_{T\bar T}(t)
=
\ell_\lambda\,
\Bigl[\mathcal O_{T\bar T}\Bigr]_{\Sigma}(t),
\label{eq:B_equals_ell_times_restriction}
\end{equation}
where the proportionality constant has been absorbed into the definition of
$\ell_\lambda$ (equivalently, into the choice of profile $\rho$).
Combining \eqref{eq:B_equals_ell_times_restriction} with
\eqref{eq:OTTbar_boundary_restricted}, we obtain
\begin{equation}
\mathcal B_{T\bar T}(t)
=
-\ell_\lambda\,
\frac{\mathcal D^2(t)}{1-\pi\lambda\,\mathcal D(t)}.
\label{eq:B_final}
\end{equation}

Substituting \eqref{eq:B_final} into \eqref{eq:OTTbar_integrated_split} and the
defining flow equation \eqref{eq:flow_global}, we finally obtain the
boundary-localized contribution to the $T\bar T$ flow,
\begin{equation}
\left(\frac{\partial W_\lambda}{\partial\lambda}\right)_\Sigma
=
-\int_\Sigma dt\;
\ell_\lambda\,
\frac{\mathcal D^2}{1-\pi\lambda\,\mathcal D}.
\label{eq:boundary_flow}
\end{equation}
By construction, \eqref{eq:boundary_flow} has dimension $L^{-2}$ and is therefore
fully consistent with \eqref{eq:flow_global}.

Equation \eqref{eq:boundary_flow} is the central result of this section.
It shows that, in a BCFT, the $T\bar T$ deformation induces a purely
one-dimensional flow supported on the physical boundary, governed entirely by
the displacement operator.
The appearance of the boundary-layer thickness $\ell_\lambda$ reflects only the
intrinsic length scale carried by the deformation parameter $\lambda$ and does
not introduce any additional running coupling or boundary degree of freedom.
The result follows solely from Ward identities, the intrinsic definition of the
$T\bar T$ deformation, and the universal BCFT boundary condition
\cite{Cardy:2018sdv,Datta:2018thy}.

\subsection{Why the boundary-localized $T\bar T$ flow is governed by the displacement operator}
\label{subsec:why_flow_is_D}

A potential source of confusion is the following.
The displacement operator $\mathcal D$ is defined already in the undeformed BCFT
by diffeomorphism Ward identities and therefore exists independently of the
$T\bar T$ deformation.
How, then, can the boundary-localized $T\bar T$ flow be expressed entirely in
terms of $\mathcal D$?

The resolution is structural rather than dynamical.
The $T\bar T$ deformation does not generate new boundary operators.
Instead, it modifies how the theory responds to geometric variations.
On a manifold with a boundary, the response to normal deformations of the
boundary embedding is governed by a single operator: the displacement operator.
As a result, any intrinsic and diffeomorphism-invariant realization of the
$T\bar T$ deformation must reduce on the boundary to a functional of
$\mathcal D$ and of the deformation parameter $\lambda$.

To make this statement precise, consider a two-dimensional quantum field theory
defined on a spacetime manifold $M$ with a timelike boundary
$\Sigma=\partial M$.
Let $y^a$ denote intrinsic coordinates on the boundary worldline, and let
$X^\mu(y)$ denote the embedding of $\Sigma$ into spacetime.
An infinitesimal deformation of the boundary shape corresponds to a normal
displacement of the embedding,
\begin{equation}
X^\mu(y)\ \longrightarrow\ X^\mu(y)+\delta x(y)\,n^\mu ,
\label{eq:shape_def_general}
\end{equation}
where $n^\mu$ is the outward-pointing unit normal to $\Sigma$.

Diffeomorphism invariance in the presence of a boundary fixes the response of the
generating functional $W$ to such a deformation.
One finds the universal relation
\begin{equation}
\delta W\big|_{\rm shape}
=
\int_\Sigma d y\; \mathcal D(y)\,\delta x(y) ,
\label{eq:shape_variation_def_D_again}
\end{equation}
which defines the displacement operator $\mathcal D$ as the operator conjugate to
normal deformations of the boundary embedding.
This definition is intrinsic and holds already in the undeformed BCFT,
independently of any irrelevant deformation.

In flat space with a straight boundary, $\mathcal D$ may be identified with the
normal--normal component of the bulk stress tensor evaluated at the boundary,
\begin{equation}
\mathcal D
=
T_{\mu\nu}\,n^\mu n^\nu\big|_{\Sigma} ,
\label{eq:D_equals_Tnn}
\end{equation}
up to improvement terms fixed by the Ward identities.
The explicit microscopic realization of $\mathcal D$ will not be important in
the following.
What matters is the conjugacy relation
\eqref{eq:shape_variation_def_D_again}, which characterizes $\mathcal D$ as the
unique operator controlling the response to boundary shape deformations.

We now turn to the intrinsic $T\bar T$ deformation.
By construction, the deformation defines a one-parameter family of theories
labelled by $\lambda$, with generating functional $W_\lambda$.
The deformation preserves locality and diffeomorphism invariance and does not
introduce new degrees of freedom.
Its defining feature is a local modification of the response to geometric
variations, equivalently a local $\lambda$-flow equation for $W_\lambda$.

Restricting attention to the boundary, locality and diffeomorphism invariance
severely constrain which operators may appear in the boundary reduction of this
flow.
Since the intrinsic boundary condition is held fixed, the only remaining
geometric degree of freedom at the boundary is a normal displacement of the
embedding \eqref{eq:shape_def_general}.
Any deformation whose effect can be reinterpreted as a change of boundary
geometry must therefore couple to the operator that generates the response to
such a displacement.

But the response to a normal deformation of the boundary embedding is, by
definition, governed by the displacement operator $\mathcal D$.
Among all local boundary operators, $\mathcal D$ is thus singled out as the
unique channel through which the $T\bar T$ deformation can act on the boundary.
Consequently, the boundary-localized $T\bar T$ flow can depend on boundary
operators only through $\mathcal D$, with any additional dependence entering
explicitly through the deformation parameter $\lambda$.

This conclusion does not rely on any particular representation of the
deformation.
Rather, it follows directly from locality, diffeomorphism invariance, and the
intrinsic definition of the displacement operator.
In the following subsections, we show how this structural fact admits two
equivalent realizations: one in which the deformation is encoded geometrically
as a shift of the boundary embedding, and another in which it is represented by
an explicit boundary functional in a fixed-boundary description.

\subsection{Fixed--moving boundary equivalence}

The relation between the displacement operator and the induced boundary shift is
an intrinsic structural property of a BCFT with a boundary.
It follows from diffeomorphism Ward identities together with the integrability of
the boundary--localized $T\bar T$ deformation, and does not rely on the
introduction of any auxiliary boundary degrees of freedom.
In particular, the equivalence discussed in this subsection concerns the
parametrization of the variational problem and the resulting space of classical
solutions, rather than the physical dynamics of an independent boundary degree
of freedom.

We begin with the response of the generating functional to deformations of the
boundary embedding.
Let $W_\lambda[x]$ denote the generating functional of the
$T\bar T$--deformed theory in the presence of a boundary embedding $x(t)$.
An infinitesimal normal deformation of the boundary,
\begin{equation}
x(t)\;\longrightarrow\; x(t)+\delta x(t),
\end{equation}
induces the universal variation
\begin{equation}
\delta W_\lambda\big|_{\rm shape}
=
\int_\Sigma dt\; \mathcal D(t)\,\delta x(t),
\label{eq:shape_variation_again}
\end{equation}
or equivalently
\begin{equation}
\frac{\delta W_\lambda}{\delta x(t)}
=
\mathcal D(t).
\label{eq:D_def_again}
\end{equation}
This Ward identity defines the displacement density $\mathcal D(t)$ as the
operator conjugate to normal deformations of the boundary embedding.
The definition is intrinsic and does not involve the deformation parameter
$\lambda$.
Any $\lambda$--dependence of observables involving $\mathcal D$ arises solely
through the state or background geometry.

A second ingredient is the intrinsic form of the boundary--localized
$T\bar T$ flow at fixed boundary embedding.
As established in the previous subsection, the deformation parameter $\lambda$
acts on the generating functional according to
\begin{equation}
\left(\frac{\partial W_\lambda}{\partial\lambda}\right)_{x}
=
\int_\Sigma dt\;
\mathcal L\!\left(\mathcal D(t),\lambda\right),
\label{eq:boundary_flow_local_again}
\end{equation}
with the boundary flow density
\begin{equation}
\mathcal L(\mathcal D,\lambda)
=
-\,\ell_\lambda\,
\frac{\mathcal D^2}{1-\pi\lambda\,\mathcal D}.
\label{eq:L_def_with_l}
\end{equation}
Here $\ell_\lambda$ denotes the intrinsic length scale generated by the
$T\bar T$ deformation, with $\ell_\lambda\propto\sqrt{\lambda}$.
The appearance of $\ell_\lambda$ ensures that
$\mathcal L$ has scaling dimension three, so that
$\int_\Sigma dt\,\mathcal L$ has the correct dimension to generate the
$\lambda$--flow of $W_\lambda$.
The flow is ultralocal along the boundary worldline and depends only on the
pre--existing displacement operator, reflecting the absence of new boundary
degrees of freedom.

The coexistence of the Ward identity \eqref{eq:shape_variation_again} and the
boundary flow equation \eqref{eq:boundary_flow_local_again} admits two equivalent
descriptions of the same intrinsic deformation.

In the moving--boundary representation, the intrinsic boundary condition is held
fixed, while the $T\bar T$ deformation is realized as a $\lambda$--dependent shift
of the boundary embedding $x(t)$.
The response of the theory to this geometric change is governed universally by
\eqref{eq:shape_variation_again}.
This description should be understood as a reparametrization of the variational
problem, rather than as the introduction of a propagating boundary degree of
freedom.

In the fixed--boundary representation, the boundary embedding is held fixed, and
the effect of the deformation is encoded entirely in the $\lambda$--dependence
of the generating functional through the boundary flow density
\eqref{eq:L_def_with_l}.
Although no explicit boundary interaction has been introduced, the structure of
the flow shows that the deformation acts only through boundary data associated
with the displacement operator.

To make the equivalence between these two representations precise, it is useful
to treat the displacement density as a classical variable and perform a
Legendre transformation.
We therefore introduce
\begin{equation}
\Gamma_\lambda[\mathcal D]
\equiv
W_\lambda[x]
-
\int_\Sigma dt\;\mathcal D(t)\,x(t),
\label{eq:Legendre_again}
\end{equation}
where $x(t)$ is now regarded as a functional of $\mathcal D(t)$ determined
implicitly by \eqref{eq:D_def_again}.
From this point onward, $\mathcal D(t)$ is interpreted as the expectation value
of the displacement operator in the $\lambda$--deformed state, or equivalently
as its classical saddle.

By construction, the conjugacy relation reads
\begin{equation}
\frac{\delta \Gamma_\lambda}{\delta \mathcal D(t)}
=
-\,x(t).
\label{eq:x_from_Gamma_again}
\end{equation}

Differentiating \eqref{eq:Legendre_again} with respect to $\lambda$ at fixed
$\mathcal D$, the $\lambda$--flow of $\Gamma_\lambda$ closes:
\begin{equation}
\left(\frac{\partial \Gamma_\lambda}{\partial\lambda}\right)_{\mathcal D}
=
\int_\Sigma dt\;
\mathcal L\!\left(\mathcal D(t),\lambda\right).
\label{eq:Gamma_flow_again}
\end{equation}
This relation expresses the integrability of the boundary--localized flow in the
space of functionals parameterized by $\mathcal D(t)$.

Integrating \eqref{eq:Gamma_flow_again} from $0$ to $\lambda$ at fixed
$\mathcal D$ yields
\begin{equation}
\Gamma_\lambda[\mathcal D]
=
\Gamma_0[\mathcal D]
+
\int_0^\lambda d\lambda'
\int_\Sigma dt\;
\mathcal L\!\left(\mathcal D(t),\lambda'\right).
\label{eq:Gamma_integrated_again}
\end{equation}
The corresponding boundary embedding follows from
\eqref{eq:x_from_Gamma_again}.
Subtracting the undeformed embedding $x_0(t)$, we define the induced boundary
shift
\begin{equation}
\delta x(t)
\equiv
x_\lambda(t)-x_0(t)
=
-\,\frac{\delta}{\delta\mathcal D(t)}
\left[
\int_0^\lambda d\lambda'\;
\mathcal L\!\left(\mathcal D(t),\lambda'\right)
\right].
\label{eq:dx_general_again}
\end{equation}

Evaluating the $\lambda'$ integral explicitly, one obtains
\begin{equation}
\delta x(t)
=
\frac{\ell_\lambda}{\pi}
\left[
\ln\!\left(1-\pi\lambda\,\mathcal D(t)\right)
+
\frac{\pi\lambda\,\mathcal D(t)}{1-\pi\lambda\,\mathcal D(t)}
\right],
\label{eq:x_shift_explicit}
\end{equation}
which holds for the expectation value (or classical saddle) of the displacement
density at fixed $\lambda$.

Equation \eqref{eq:x_shift_explicit} makes the equivalence between the fixed--
boundary and moving--boundary representations explicit.
The same intrinsic boundary--localized $T\bar T$ deformation may be described
either as a $\lambda$--dependent reparametrization of the boundary embedding or
as a modification of the variational principle at fixed boundary.
The two descriptions differ only by a choice of variables and provide equivalent
parameterizations of the same boundary physics.

\subsection{Integration of the flow and induced boundary action}

In the previous subsection we established that the boundary--localized
$T\bar T$ deformation admits two equivalent descriptions:
it may be interpreted either as a $\lambda$--dependent reparametrization of the
boundary embedding or, equivalently, as a deformation of the variational
principle at fixed boundary.
This equivalence follows directly from diffeomorphism Ward identities together
with the integrability of the boundary--localized flow, and does not rely on the
introduction of any additional boundary degrees of freedom.

In this subsection we work entirely within the fixed--boundary representation
and integrate the flow explicitly.
Our goal is not to promote the result to an independent boundary dynamics, but
rather to exhibit a boundary functional whose variation reproduces the same
$\lambda$--flow.
Such a functional should be regarded as a convenient parametrization of the
trajectory in theory space generated by the $T\bar T$ deformation.

Since the $T\bar T$ deformation is defined as an evolution in the coupling
$\lambda$, a natural object in the fixed--boundary description is the cumulative
effect of the boundary--supported flow along this trajectory.
We therefore define the induced boundary contribution
\begin{equation}
I_{\rm 1d}
\equiv
-\int_0^\lambda d\lambda'\,
\left(\frac{\partial W_{\lambda'}}{\partial\lambda'}\right)_{x},
\label{eq:I1d_def_refined}
\end{equation}
where the derivative is taken at fixed boundary embedding.
The overall minus sign is chosen so that $I_{\rm 1d}$ may be interpreted as an
effective boundary term added to the undeformed generating functional.

Using the boundary--localized flow derived previously, the $\lambda'$--derivative
of the generating functional takes the universal form
\begin{equation}
\left(\frac{\partial W_{\lambda'}}{\partial\lambda'}\right)_{x}
=
-\int_\Sigma dt\;
\ell_{\lambda'}\,
\frac{\mathcal D^2(t)}{1-\pi\lambda'\,\mathcal D(t)} ,
\label{eq:boundary_flow_refined}
\end{equation}
where $\mathcal D(t)$ is the displacement operator defined intrinsically by the
shape Ward identity
$\delta W|_{\rm shape}=\int_\Sigma dt\,\mathcal D\,\delta x$.
The quantity $\ell_{\lambda'}$ denotes the unique length scale generated by the
$T\bar T$ deformation and satisfies $\ell_{\lambda'}\propto\sqrt{\lambda'}$ up to
a scheme--dependent dimensionless constant.

It is important to emphasize that $\ell_{\lambda'}$ is not an independent
coupling and does not represent an additional running parameter.
Once $\lambda'$ is specified, $\ell_{\lambda'}$ is completely fixed by
dimensional analysis.
Consequently, even in the presence of a boundary, the $T\bar T$ deformation
defines a genuine one--parameter family of theories.

The displacement operator $\mathcal D(t)$ exists already in the undeformed BCFT
and is defined independently of the deformation.
In particular, $\mathcal D(t)$ itself does not undergo a $\lambda$--flow.
All $\lambda'$--dependence in \eqref{eq:boundary_flow_refined} arises from the
universal nonlinear structure dictated by the $T\bar T$ deformation.

As a result, the integration over $\lambda'$ in
\eqref{eq:I1d_def_refined} may be performed at fixed $\mathcal D$.
Substituting \eqref{eq:boundary_flow_refined} into
\eqref{eq:I1d_def_refined}, we obtain
\begin{equation}
I_{\rm 1d}
=
\int_0^\lambda d\lambda'
\int_\Sigma dt\;
\ell_{\lambda'}\,
\frac{\mathcal D^2(t)}{1-\pi\lambda'\,\mathcal D(t)} .
\label{eq:I1d_intermediate_refined}
\end{equation}

At this stage it is neither necessary nor particularly illuminating to insist
on a fully closed expression for $I_{\rm 1d}$.
Keeping $\ell_{\lambda'}$ explicit, one may integrate by parts in $\lambda'$
to obtain the equivalent representation
\begin{equation}
I_{\rm 1d}
=
-\frac{1}{\pi}
\int_\Sigma dt\;
\mathcal D(t)
\left[
\ell_\lambda\,
\ln\!\left(1-\pi\lambda\,\mathcal D(t)\right)
-
\int_0^\lambda d\lambda'\;
\frac{d\ell_{\lambda'}}{d\lambda'}\,
\ln\!\left(1-\pi\lambda'\,\mathcal D(t)\right)
\right].
\label{eq:I1d_general_refined}
\end{equation}
The second term reflects the scheme dependence associated with the precise
normalization of $\ell_{\lambda'}$.
Different choices of scheme correspond to adding local boundary counterterms and
do not affect the intrinsic structure of the flow.

Equation \eqref{eq:I1d_intermediate_refined}, together with the equivalent form
\eqref{eq:I1d_general_refined}, is sufficient for our purposes.
It shows that the integrated effect of the $T\bar T$ deformation can be encoded
in a boundary--localized functional built entirely from the pre--existing
displacement operator.
No new boundary degrees of freedom are introduced, and the appearance of the
length scale $\ell_\lambda$ reflects only the intrinsic scale carried by the
deformation parameter $\lambda$, not the emergence of an additional coupling.

\subsection{Inverting the boundary shift to obtain $\mathcal D$ and holographic specialization}
\label{subsec:invert_shift_D}

In the boundary--localized $T\bar T$ flow derived above, the deformation is
governed by a single scalar boundary operator, the displacement density
$\mathcal D(t)$.
This operator is defined intrinsically by diffeomorphism Ward identities and does
not itself undergo any $\lambda$--evolution.
Instead, the effect of the $T\bar T$ deformation is encoded in how physical
states, expectation values of observables, and geometric data depend on the
deformation parameter $\lambda$.

Throughout this subsection, $\mathcal D(t)$ denotes the expectation value (or
classical saddle) of the displacement operator in the $\lambda$--deformed state.
The operator itself is $\lambda$--independent and is fixed once and for all by
the Ward identity conjugate to boundary shape deformations.
In particular, $\mathcal D$ should be regarded as the quantity conjugate to
infinitesimal normal deformations of the boundary embedding.

A nonvanishing displacement density modifies the location of the boundary
worldline, while the induced boundary shift $\delta x$ is a directly measurable
geometric quantity.
In the fixed--boundary representation, $\mathcal D$ is treated as a fundamental
boundary variable, whereas in the moving--boundary representation the same
information is encoded geometrically in the $\lambda$--dependent embedding
$x(t)$.
On the perturbative branch continuously connected to the undeformed BCFT, these
two descriptions are related by a Legendre transform and are therefore
equivalent.


At fixed $\lambda$, the exact relation between the boundary shift and the
displacement density takes the form
\begin{equation}
\delta x(t)
=
\frac{\ell_\lambda}{\pi}
\left[
\ln\!\left(1-\pi\lambda\,\mathcal D(t)\right)
+
\frac{\pi\lambda\,\mathcal D(t)}{1-\pi\lambda\,\mathcal D(t)}
\right],
\label{eq:x_shift_explicit_app_refined}
\end{equation}
where $\ell_\lambda\propto\sqrt{\lambda}$ is the intrinsic length scale generated
by the $T\bar T$ deformation.
The appearance of $\ell_\lambda$ restores the correct dimension $[\delta x]=L$
and reflects the fact that the relation is purely kinematical at fixed $\lambda$.

Introducing the dimensionless variables
\begin{equation}
y(t)\equiv \frac{\pi\,\delta x(t)}{\ell_\lambda},
\qquad
u(t)\equiv 1-\pi\lambda\,\mathcal D(t),
\end{equation}
the relation may be rewritten in the algebraic form
\begin{equation}
y=\ln u+\frac{1}{u}-1 .
\end{equation}
This equation can be inverted exactly in terms of the Lambert $W$ function,
leading to
\begin{equation}
\mathcal D(t)
=
\frac{1}{\pi\lambda}
\left[
1+\frac{1}{W_k\!\left(
-\exp\!\left\{-\left[1+\frac{\pi\,\delta x(t)}{\ell_\lambda}\right]\right\}
\right)}
\right],
\label{eq:D_general_inverted_refined}
\end{equation}
where $k$ labels the branch.

For real boundary shifts $\delta x\ge0$, the argument of the Lambert function lies
in the interval $-1/e\le z<0$, on which two real branches, $W_0$ and $W_{-1}$,
exist.
Requiring continuity with the undeformed BCFT,
\begin{equation}
\delta x\to0 \quad \Longrightarrow \quad \mathcal D\to0 ,
\end{equation}
selects the $W_{-1}$ branch.
On this branch, the inversion is smooth for sufficiently small
$\pi\lambda\,\mathcal D$, corresponding to the perturbative
(\emph{soft--boundary}) regime in which the Legendre transform relating the
fixed-- and moving--boundary descriptions is invertible.


We now specialize to the static holographic realization.
As will be shown in the following section, the boundary shift is time independent
and is determined geometrically from the bulk solution as
\begin{equation}
\Delta x_{\mathrm{bdy}}
=
z_c\,\sinh\!\left(\frac{\rho_0}{\ell}\right),
\end{equation}
while the deformation parameter is related to the cutoff position by
\begin{equation}
\lambda=\frac{8G_N}{\ell}\,z_c^2 .
\end{equation}
Eliminating $z_c$ yields
\begin{equation}
\delta x(\lambda)
=
\sqrt{\frac{\lambda\,\ell}{8G_N}}\;
\sinh\!\left(\frac{\rho_0}{\ell}\right).
\end{equation}

Substituting the holographic displacement $\delta x(\lambda)$ into the
\emph{inverted} formula \eqref{eq:D_general_inverted_refined} would implicitly
assume that the saddle remains on the perturbative branch selected by the
continuity condition $\delta x\!\to\!0 \Rightarrow \mathcal D\!\to\!0$, i.e.\ the
$W_{-1}$ branch.
As emphasized above, the static holographic saddle is \emph{not} continuously
connected to the undeformed BCFT along this soft--boundary branch, and therefore
one should not diagnose its scaling by directly plugging into the
$W_{-1}$ inversion.
Instead, the correct conclusion about the holographic scaling follows already
from the \emph{direct} relation \eqref{eq:x_shift_explicit_app_refined}, viewed
as an implicit equation for $\mathcal D$ at fixed $\lambda$.

Starting from \eqref{eq:x_shift_explicit_app_refined}, define the dimensionless
combination
\begin{equation}
a(t)\equiv \pi\lambda\,\mathcal D(t),
\qquad
y(t)\equiv \frac{\pi\,\delta x(t)}{\ell_\lambda},
\end{equation}
so that \eqref{eq:x_shift_explicit_app_refined} becomes the purely algebraic
relation
\begin{equation}
y
=
\ln(1-a)+\frac{a}{1-a}.
\label{eq:y_of_a}
\end{equation}
Crucially, at fixed $\lambda$ the right-hand side depends only on the single
variable $a$, and is independent of $\lambda$.
Therefore, for any given value of the ratio $y=\pi\delta x/\ell_\lambda$, the
equation \eqref{eq:y_of_a} fixes $a$ to a \emph{number} $a=a(y)$ (possibly with
multiple branches), and one has identically
\begin{equation}
\mathcal D(t)=\frac{a(y(t))}{\pi\lambda}.
\label{eq:D_scaling_general}
\end{equation}
In particular, the $\lambda$--scaling of $\mathcal D$ is entirely controlled by
the $\lambda$--scaling of the dimensionless ratio $y=\pi\delta x/\ell_\lambda$.

In the static holographic solution,
\begin{equation}
\delta x(\lambda)=\sqrt{\frac{\lambda\,\ell}{8G_N}}\,
\sinh\!\left(\frac{\rho_0}{\ell}\right),
\qquad
\ell_\lambda \propto \sqrt{\lambda},
\end{equation}
so that
\begin{equation}
y(\lambda)=\frac{\pi\,\delta x(\lambda)}{\ell_\lambda}
\;\longrightarrow\;
y_0\neq 0
\qquad (\lambda\to0^+),
\end{equation}
with $y_0$ a finite nonzero constant determined by the holographic data and the
scheme choice in $\ell_\lambda$.
Equation \eqref{eq:y_of_a} then fixes $a\to a_0\equiv a(y_0)$ with $a_0$ an
$\mathcal O(1)$ constant (again, possibly on a nonperturbative branch), and
\eqref{eq:D_scaling_general} immediately implies
\begin{equation}
\mathcal D(\lambda)\sim \frac{a_0}{\pi}\,\frac{1}{\lambda},
\qquad
\lambda\to0^+.
\label{eq:D_diverge_holo}
\end{equation}
Thus the divergence $\mathcal D\sim 1/\lambda$ is not an artifact of a particular
Lambert--$W$ branch choice in the inversion formula; it follows directly from
the implicit relation \eqref{eq:x_shift_explicit_app_refined} together with the
holographic scaling $\delta x\sim \ell_\lambda\sim \sqrt{\lambda}$.

The physical interpretation is that the holographic saddle approaches a
rigid--boundary regime: the geometric shift $\delta x$ vanishes with the cutoff,
yet the conjugate displacement density diverges in such a way that the
dimensionless combination $a=\pi\lambda\mathcal D$ remains finite.
Equivalently, the boundary behaves as an infinitely heavy object whose position
is pinned by a diverging conjugate stress, placing the holographic saddle
outside the soft--boundary branch where the fixed-- and moving--boundary
representations are Legendre--equivalent.

This divergence does not signal an inconsistency of the formalism.
Rather, it indicates that the holographic saddle lies outside the perturbative
soft--boundary branch on which the fixed-- and moving--boundary descriptions are
Legendre--equivalent.
Physically, $\mathcal D$ measures the conjugate stress required to deform the
boundary position.
The limit $\delta x\to0$ accompanied by $\mathcal D\to\infty$ therefore describes
a \emph{rigid or infinitely heavy boundary}, whose position is pinned by an
infinitely large conjugate stress rather than determined by a smooth response.

In this rigid--boundary regime, the moving--boundary (geometric) description
remains perfectly well defined, while the fixed--boundary representation in
terms of a finite displacement density breaks down.
The holographic realization thus naturally selects a nonperturbative branch in
which the boundary position is enforced as a hard constraint, rather than
generated by a soft $T\bar T$ deformation continuously connected to the
undeformed BCFT.

\section{Type~A: boundary deformed}
\label{sec2}

In this section we discuss a holographic realization of the boundary--localized
$T\bar T$ deformation in which the physical boundary of the BCFT is displaced by
a finite amount.
Following standard terminology in the AdS$_3$/BCFT literature, we refer to this
realization as \emph{Type~A}.
This classification is not introduced here for the first time ~\cite{Wang:2024jem}; rather, it
provides a convenient label for a well--studied geometric setup whose role in the
intrinsic description of boundary--localized $T\bar T$ deformations we now
clarify.

The Type~A realization is characterized by a genuine deformation of the physical
BCFT boundary.
In this setup, the boundary embedding is shifted by a finite amount along the
cutoff surface, and the associated displacement operator acquires a nonzero
expectation value.
As we will see, this boundary motion is not a matter of coordinate choice or
parametrization, but arises unavoidably from the bulk geometry once the holographic
construction is specified.

Concretely, Type~A corresponds to a holographic implementation in which the bulk
geometry is truncated by a rigid Dirichlet cutoff surface placed at fixed radial
position, while the end--of--the--world (EOW) brane $Q$ continues to satisfy its
standard Neumann boundary condition with constant tension.
The cutoff surface is treated as non--dynamical: its embedding is fixed once and
for all, and no mixed or Neumann boundary condition is imposed on it.
By contrast, the embedding of the EOW brane is determined dynamically by the
Neumann condition and is therefore insensitive to the introduction of the cutoff.

Because the cutoff surface is fixed while the EOW brane embedding is dynamical,
the intersection point between the cutoff surface and the brane is not anchored
to a fixed boundary location.
As the deformation parameter $\lambda$ is varied, this intersection point moves
along the direction parallel to the boundary.
When the induced geometry on the cutoff surface is interpreted as the spacetime
of the deformed BCFT, this motion manifests itself as a finite displacement of the
physical boundary.
In particular, the boundary is located at a $\lambda$--dependent position that
cannot be removed by a reparameterization of boundary coordinates.

Throughout this section we work within the standard AdS$_3$/BCFT$_2$ framework
developed in Refs.~\cite{Takayanagi:2011zk,Fujita:2011fp}.
The bulk spacetime is asymptotically AdS$_3$ and contains an EOW brane $Q$ of
constant tension.
The $T\bar T$ deformation is implemented holographically by introducing a
Dirichlet cutoff surface at fixed radial position, whose location is related to
the deformation parameter $\lambda$ in the usual way.

The essential point of the Type~A construction is therefore geometric.
Since the EOW brane is not tied to a fixed point on the cutoff surface, its
intersection with the cutoff surface occurs at a finite and $\lambda$--dependent
boundary location.
As a result, the induced displacement of the BCFT boundary is a genuine physical
effect.
Equivalently, the normal--normal component of the stress tensor at the boundary
does not vanish, and the displacement operator
$\mathcal D = T_{xx}\big|_{\Sigma}$ acquires a nonzero expectation value.

It is important to stress that this boundary displacement is not part of the
\emph{intrinsic} definition of the $T\bar T$ deformation.
In the intrinsic formulation developed in
Section~\ref{sec:boundaryflow}, the boundary embedding was held fixed and the
effect of the deformation was encoded in a nonlinear boundary functional built
from the displacement operator.
Type~A instead provides a holographic realization in which the same intrinsic
deformation is represented geometrically as a finite shift of the boundary
embedding.
The equivalence between these two descriptions relies precisely on the fact that
the displacement operator is nontrivial in Type~A and mediates the mapping between
boundary interactions and boundary motion.

In the following subsections we make this correspondence explicit.
We compute the induced boundary shift $\Delta x_{\rm bdy}$ directly from the bulk
geometry and show that it reproduces the field--theoretic relation between
$\Delta x_{\rm bdy}$ and the displacement operator implied by the
boundary--localized $T\bar T$ flow.

\subsection{Zero temperature case: pure AdS$_3$}

We begin with Euclidean AdS$_3$ in Poincar\'e coordinates,
\begin{equation}
ds^2=\frac{\ell^2}{z^2}\left(d\tau^2+dz^2+dx^2\right),
\label{eq:AdS3_Poincare}
\end{equation}
and recall that the same geometry admits an AdS$_2$ slicing adapted to the
BCFT boundary.
In Gaussian normal coordinates this slicing takes the form
\begin{equation}
ds^2
=
d\rho^2+\ell^2\cosh^2\!\left(\frac{\rho}{\ell}\right)\frac{d\tau^2+dy^2}{y^2},
\label{eq:AdS3_AdS2slicing}
\end{equation}
where each surface of constant $\rho$ is an AdS$_2$ space with induced metric
\begin{equation}
ds^2_{\mathrm{AdS}_2}
=
\ell^2\cosh^2\!\left(\frac{\rho}{\ell}\right)\frac{d\tau^2+dy^2}{y^2}.
\end{equation}

The end-of-the-world (EOW) brane $Q$ is determined by the Neumann boundary
condition
\begin{equation}
K^{(\gamma)}_{ab}-\gamma_{ab}K^{(\gamma)}
=
8\pi G_N\,T^Q_{ab},
\label{eq:NBC}
\end{equation}
where $\gamma_{ab}$ denotes the induced metric on $Q$.
For a constant-tension brane with $T^Q_{ab}=T\,\gamma_{ab}$, this reduces to
\begin{equation}
K^{(\gamma)}_{ab} = (K^{(\gamma)}-8\pi G_N T)\,\gamma_{ab}.
\end{equation}
In the AdS$_2$ slicing \eqref{eq:AdS3_AdS2slicing}, hypersurfaces of constant
$\rho$ have extrinsic curvature proportional to the induced metric and therefore
solve the Neumann condition.
Evaluating \eqref{eq:NBC} at $\rho=\rho_0$ fixes the tension to be
\begin{equation}
T=\frac{1}{\ell}\tanh\!\left(\frac{\rho_0}{\ell}\right),
\label{eq:T_rho0}
\end{equation}
which we use to parameterize the BCFT boundary condition.

The associated boundary entropy is
\begin{equation}
S_{\mathrm{bdy}}
=
\frac{\rho_0}{4G_N},
\label{eq:Sbdy_rho0}
\end{equation}
in agreement with both holographic entanglement entropy and thermodynamic
computations in AdS/BCFT.

The $T\bar T$ deformation is implemented holographically by introducing a
Dirichlet cutoff surface at fixed radial position
\begin{equation}
z=z_c,
\label{eq:cutoff_zc}
\end{equation}
on which the deformed theory is defined.
Matching the Brown--York stress tensor at $z=z_c$ with the intrinsic trace flow
fixes the relation between the cutoff scale and the deformation parameter to be
\begin{equation}
\lambda=\frac{8G_N}{\ell}\,z_c^2 .
\label{eq:lambda_zc}
\end{equation}

\begin{figure}
	\centering
	\includegraphics[width=12cm,height=3.7cm]{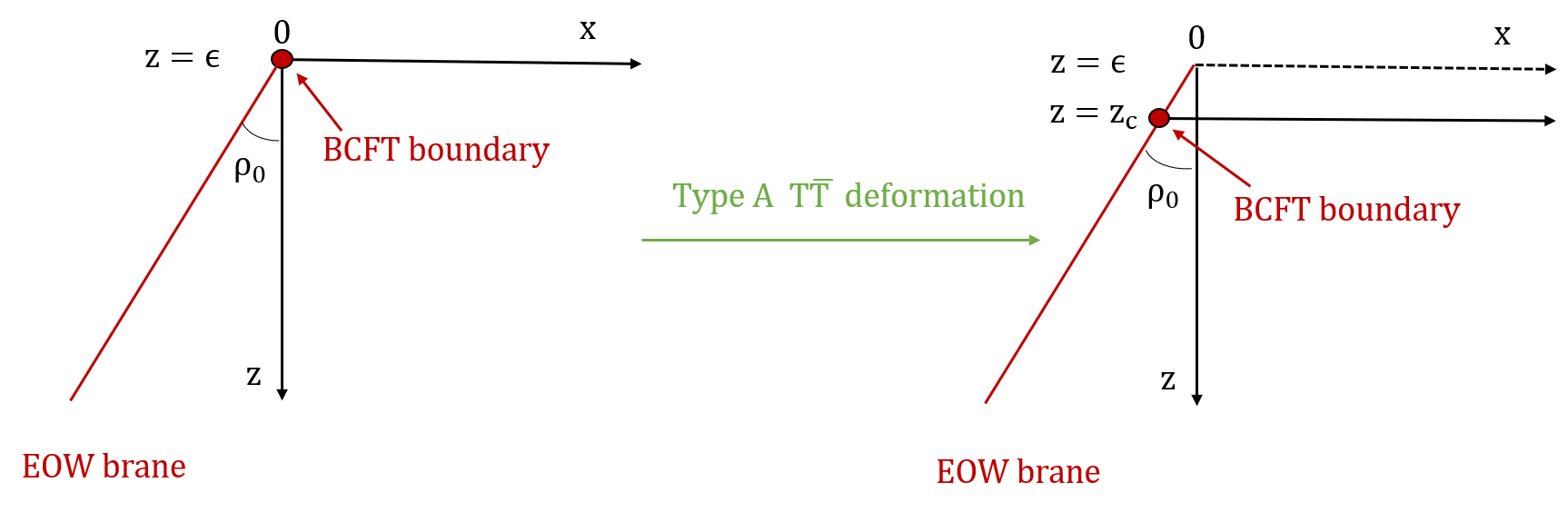}\\
	\caption{Bulk picture of the Type~A deformation in pure AdS$_3$.}
	\label{A}
\end{figure}

In the Type~A realization, the cutoff surface is kept rigidly at $z=z_c$, while
the EOW brane remains at $\rho=\rho_0$.
The physical BCFT boundary is identified with the intersection of these two
surfaces.
Geometrically, this shifts the boundary endpoint from $(x,z)=(0,0)$ in the
undeformed theory to
\begin{equation}
(x,z)=\left(-z_c\sinh\!\frac{\rho_0}{\ell},\,z_c\right),
\label{eq:TypeA_shift_pureAdS}
\end{equation}
so that the boundary location on the cutoff surface is displaced by
\begin{equation}
\Delta x_{\mathrm{bdy}}
=
z_c\sinh\!\left(\frac{\rho_0}{\ell}\right).
\end{equation}
This finite displacement of the BCFT boundary is the defining geometric feature
of the Type~A picture.
The bulk configuration is shown schematically in Fig.~\ref{A}.

We now compute the entanglement entropy of an interval $A$ of coordinate length
$L$ on the cutoff surface.
Because the BCFT boundary is shifted, it is convenient to parameterize the
interval as
\begin{equation}
A:\quad
x\in\left[-z_c\sinh\!\left(\frac{\rho_0}{\ell}\right),\;
L-z_c\sinh\!\left(\frac{\rho_0}{\ell}\right)\right],
\label{eq:intervalA_TypeA_pureAdS}
\end{equation}
so that its left endpoint coincides with the displaced boundary.

\begin{figure}
	\centering
	\includegraphics[width=5.5cm,height=4cm]{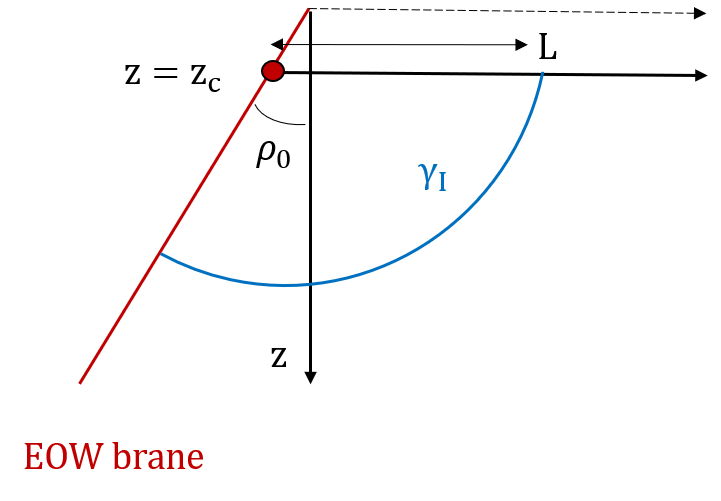}\\
	\caption{RT surface for the Type~A deformation in pure AdS$_3$.}
	\label{ART}
\end{figure}

On the gravity side, the entanglement entropy is computed using the
Ryu--Takayanagi prescription.
The relevant extremal surface is a geodesic in the truncated wedge geometry,
which can be obtained by reflecting the geodesic across the EOW brane.
Evaluating the regulated geodesic length between the endpoints on the cutoff
surface yields
\begin{equation}
\begin{split}
S(A)
&=\frac{\mathrm{Area}(\gamma_A)}{4G_N}
\\
&=
\frac{\ell}{4G_N}\log\!\left(\frac{2L}{z_c}\right)
+\frac{\rho_0}{4G_N}
-\frac{z_c\ell\sinh\!\left(\frac{\rho_0}{\ell}\right)}{4LG_N}
+\mathcal{O}(z_c^2).
\end{split}
\label{eq:TypeA_EE_gravity_pureAdS}
\end{equation}
The corresponding RT configuration is shown in Fig.~\ref{ART}.

From the field-theory perspective, two effects contribute to the entanglement
entropy.
First, the UV cutoff $\epsilon$ of the undeformed BCFT is replaced by the finite
cutoff $z_c$.
Second, the physical boundary is shifted by $\Delta x_{\mathrm{bdy}}$.
The portion of the interval between
$x=-z_c\sinh(\rho_0/\ell)$ and $x=0$ lies in the wedge region between
$\rho=\rho_0$ and $\rho=0$ and contributes precisely the boundary entropy
$S_{\mathrm{bdy}}$.
The remaining segment has effective length
\begin{equation}
L' = L - z_c\sinh\!\left(\frac{\rho_0}{\ell}\right).
\end{equation}
Using the standard ground-state entanglement entropy formula for a BCFT interval,
we obtain
\begin{equation}
S(A)
=
\frac{c}{6}\log\!\left(\frac{2L'}{z_c}\right)
+
S_{\mathrm{bdy}} .
\end{equation}
Expanding for small $z_c$ gives
\begin{equation}
\begin{split}
S(A)
&=
\frac{c}{6}\log\!\left(\frac{2L}{z_c}\right)
+S_{\mathrm{bdy}}
-\frac{c\,z_c\sinh(\rho_0/\ell)}{6L}
+\mathcal{O}(z_c^2),
\end{split}
\label{eq:TypeA_EE_field_pureAdS}
\end{equation}
which agrees precisely with the holographic result
\eqref{eq:TypeA_EE_gravity_pureAdS} upon using
$c=\frac{3\ell}{2G_N}$ together with \eqref{eq:Sbdy_rho0}.

\subsection{Finite temperature case: BTZ}

We now turn to the finite-temperature case, for which the bulk geometry is the
Euclidean BTZ black hole,
\begin{equation}
ds^2
=
\ell^2\left(
\frac{f(z)}{z^2}d\tau^2+\frac{dz^2}{f(z)z^2}+\frac{dx^2}{z^2}
\right),
\qquad
f(z)=1-\left(\frac{z}{z_H}\right)^2 ,
\label{eq:btz_metric}
\end{equation}
with Euclidean time periodically identified as $\tau\sim\tau+2\pi z_H$.
The dual BCFT is therefore at finite temperature
\begin{equation}
T_{\mathrm{BCFT}}=\frac{1}{2\pi z_H},
\qquad
\beta=2\pi z_H .
\end{equation}

As in the zero-temperature case, the end-of-the-world brane $Q$ is determined by
the Neumann boundary condition \eqref{eq:NBC} with constant tension.
Solving this condition in the BTZ background yields the brane trajectory
\begin{equation}
x_Q(z)
=
-\,z_H\,\mathrm{arcsinh}\!\left(
\frac{\ell T\,z}{z_H\sqrt{1-\ell^2T^2}}
\right),
\label{eq:TypeA_Q_BTZ}
\end{equation}
which smoothly reduces to the pure AdS profile in the limit $z_H\to\infty$, as
required.

\begin{figure}
	\centering
	\includegraphics[width=13cm,height=3.7cm]{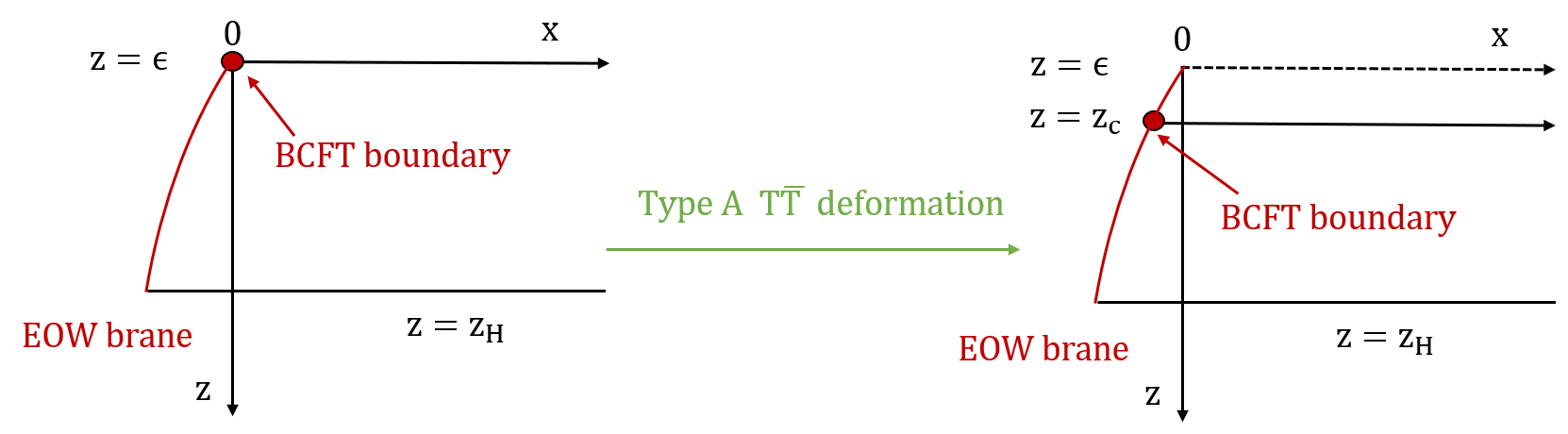}\\
	\caption{Bulk picture of the Type~A deformation in the BTZ background.}
	\label{A1}
\end{figure}

The $T\bar T$ deformation is again implemented by introducing a Dirichlet cutoff
surface at fixed radial position $z=z_c$.
In the Type~A realization, the EOW brane remains fixed at the profile
\eqref{eq:TypeA_Q_BTZ}, while the physical BCFT boundary is identified with the
intersection of the cutoff surface and the brane.
This intersection point is located at
\begin{equation}
(x,z)=\left(x_Q(z_c),\,z_c\right)
=
\left(
-\,z_H\,\mathrm{arcsinh}\!\left(
\frac{\ell T\,z_c}{z_H\sqrt{1-\ell^2T^2}}
\right),
\,z_c
\right),
\label{eq:TypeA_shift_BTZ}
\end{equation}
so that the boundary on the cutoff surface is displaced by
\begin{equation}
\Delta x_{\mathrm{bdy}}
=
-\,x_Q(z_c)
=
z_H\,\mathrm{arcsinh}\!\left(
\frac{\ell T\,z_c}{z_H\sqrt{1-\ell^2T^2}}
\right).
\end{equation}
This finite displacement generalizes the zero-temperature result and encodes the
combined effect of the $T\bar T$ deformation and finite temperature.
The corresponding bulk configuration is illustrated in Fig.~\ref{A1}.

We now compute the entanglement entropy of an interval $A$ of coordinate length
$L$ on the cutoff surface, anchored at the shifted boundary,
\begin{equation}
A:\quad
x\in\left[x_Q(z_c),\,x_Q(z_c)+L\right].
\end{equation}
It is convenient to introduce the effective interval length
\begin{equation}
L'
=
L-\Delta x_{\mathrm{bdy}},
\label{eq:TypeA_Lprime_BTZ}
\end{equation}
which measures the portion of the interval extending into the dynamical BCFT
region beyond the wedge associated with the EOW brane.

\begin{figure}
	\centering
	\includegraphics[width=6.5cm,height=4cm]{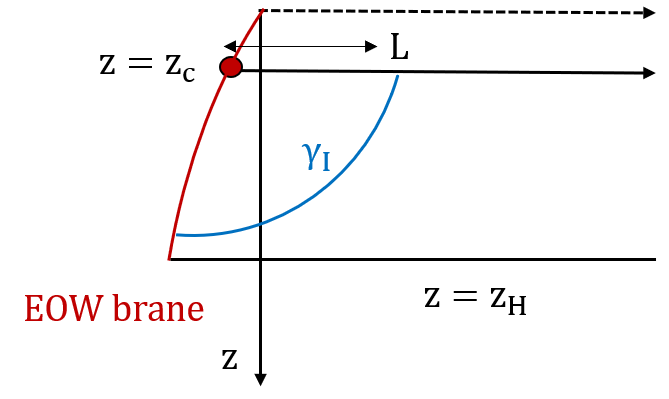}\\
	\caption{RT surface for the Type~A deformation in the BTZ case.}
	\label{A1RT}
\end{figure}

On the gravity side, the entanglement entropy is computed using the
Ryu--Takayanagi prescription.
The relevant extremal surface is a BTZ geodesic in the truncated wedge geometry.
Expanding the regulated geodesic length at small cutoff $z_c$, one finds
\begin{equation}
S(A)
=
\frac{c}{6}\log\!\left(\frac{\beta}{\pi z_c}
\sinh\!\frac{2\pi L}{\beta}\right)
+S_{\mathrm{bdy}}
-\frac{\pi c\,\ell T\,z_c}{3\beta\sqrt{1-\ell^2T^2}}
\coth\!\left(\frac{2\pi L}{\beta}\right)
+\mathcal{O}(z_c^2),
\label{eq:TypeA_EE_gravity_BTZ_expand}
\end{equation}
where $S_{\mathrm{bdy}}=\rho_0/(4G_N)$ as in the zero-temperature case.
The corresponding extremal surface is shown in Fig.~\ref{A1RT}.

From the field-theory perspective, the entanglement entropy is given by the
standard thermal BCFT formula evaluated on an interval of effective length $L'$
with UV cutoff $z_c$,
\begin{equation}
S_{\mathrm{BCFT}}(L')
=
\frac{c}{6}\log\!\left(\frac{\beta}{\pi z_c}
\sinh\!\frac{2\pi L'}{\beta}\right)
+S_{\mathrm{bdy}} .
\label{eq:thermal_BCFT_EE}
\end{equation}
Substituting \eqref{eq:TypeA_Lprime_BTZ} and expanding for small $z_c$ yields
\begin{equation}
\begin{split}
S(A)
&=
\frac{c}{6}\log\!\left(\frac{\beta}{\pi z_c}
\sinh\!\frac{2\pi L}{\beta}\right)
+S_{\mathrm{bdy}}
-\frac{\pi c}{3\beta}\,\Delta x_{\mathrm{bdy}}\,
\coth\!\left(\frac{2\pi L}{\beta}\right)
+\mathcal{O}(z_c^2)
\\
&=
\frac{c}{6}\log\!\left(\frac{\beta}{\pi z_c}
\sinh\!\frac{2\pi L}{\beta}\right)
+S_{\mathrm{bdy}}
-\frac{\pi c\,\ell T\,z_c}{3\beta\sqrt{1-\ell^2T^2}}
\coth\!\left(\frac{2\pi L}{\beta}\right)
+\mathcal{O}(z_c^2),
\end{split}
\end{equation}
which is in precise agreement with the gravitational result
\eqref{eq:TypeA_EE_gravity_BTZ_expand}.

\section{Type B: boundary preserved}
\label{sec3}

In this section we present a second holographic realization of the
boundary-localized $T\bar T$ deformation derived in
Section~\ref{sec:boundaryflow}, which we refer to as \emph{Type~B} ~\cite{Wang:2024jem}.
In contrast to Type~A, this realization does not give rise to a finite
deformation of the physical BCFT boundary.
The boundary remains fixed in the induced field-theory geometry, and no
explicit boundary displacement is generated.
Crucially, this property is neither imposed by hand nor a consequence of
choosing a particular field-theoretic representation.
Rather, it follows directly from the geometric structure of the bulk cutoff
construction itself.

The defining feature of the Type~B realization is the choice of Dirichlet
cutoff surface.
Instead of placing the cutoff at a surface of fixed radial position, the
cutoff is chosen to follow a nontrivial trajectory in the bulk such that its
induced metric is asymptotically $\mathrm{AdS}_2$.
With this choice, the cutoff surface intersects the end-of-the-world (EOW)
brane only at asymptotic infinity.
As a result, the physical BCFT boundary is identified with the conformal
boundary of an $\mathrm{AdS}_2$ geometry.
Because this boundary is asymptotic rather than located at a finite
intersection point, it is geometrically pinned and admits no finite normal
displacement under variations of the deformation parameter.

The absence of boundary motion in Type~B therefore has a purely geometric
origin.
Once the cutoff surface is required to be asymptotically $\mathrm{AdS}_2$,
there exists no finite boundary location whose embedding could be shifted.
The preservation of the BCFT boundary is thus fixed already at the level of
the bulk construction, prior to any discussion of boundary dynamics or
operator content.

From the field-theory perspective, this geometric pinning has an immediate
and concrete consequence.
Since the BCFT boundary coincides with the conformal boundary of
$\mathrm{AdS}_2$, the normal--normal component of the stress tensor evaluated
at the boundary vanishes,
\begin{equation}
T_{nn}\big|_{\Sigma}=0,
\end{equation}
and hence the displacement operator,
\begin{equation}
\mathcal D \equiv T_{nn}\big|_{\Sigma},
\end{equation}
is identically zero.
It is important to emphasize that this vanishing should be understood as a
\emph{result} of the geometric setup rather than as its cause.
In particular, the absence of a nontrivial displacement operator does not
signal the absence of a $T\bar T$ deformation.
Instead, it reflects the fact that, in this realization, the displacement
channel is geometrically frozen and the intrinsic boundary-localized flow is
absorbed entirely into the background geometry of the cutoff surface.

Throughout this section we work within the standard
$\mathrm{AdS}_3/\mathrm{BCFT}_2$ framework introduced in
Refs.~\cite{Takayanagi:2011zk,Fujita:2011fp}, with an EOW brane $Q$ of constant
tension satisfying Neumann boundary conditions.
The distinction between Type~A and Type~B lies entirely in the choice of
cutoff geometry.
While Type~A admits a finite intersection between the cutoff surface and the
EOW brane, leading to a nonvanishing displacement operator and an explicit
boundary shift, Type~B enforces an asymptotically $\mathrm{AdS}_2$ cutoff that
removes this possibility altogether.

As a consequence, the $T\bar T$ deformation in the Type~B realization is
encoded entirely through the bulk cutoff geometry and its effect on bulk
observables, while the physical BCFT boundary itself remains fixed.
In the following subsections we construct the Type~B geometry explicitly,
first for pure $\mathrm{AdS}_3$ and then for the BTZ black hole, and verify
that physical observables such as entanglement entropy agree with those
obtained in the Type~A realization.
This agreement provides a nontrivial consistency check of the intrinsic
equivalence between the two holographic realizations, despite their sharply
different geometric implementations.

\subsection{Zero temperature case: pure AdS$_3$}

We work in the AdS$_2$ slicing coordinates \eqref{eq:AdS3_AdS2slicing}.
In the Type~B realization, the bulk region is defined as a finite slab bounded by
two constant-$\rho$ hypersurfaces,
\begin{equation}
ds^2
=
d\rho^2+\ell^2\cosh^2\!\left(\frac{\rho}{\ell}\right)\frac{d\tau^2+dy^2}{y^2},
\qquad
\rho_0 \le \rho \le \rho_c .
\label{eq:TypeB_slab_pureAdS}
\end{equation}
The hypersurface at $\rho=\rho_0$ is the end-of-the-world brane $Q$, which obeys
the Neumann boundary condition with constant tension.
As reviewed previously, the brane position $\rho_0$ is fixed by the tension
through \eqref{eq:T_rho0}, and the associated boundary entropy is given by
\eqref{eq:Sbdy_rho0}.

The hypersurface at $\rho=\rho_c$ implements the holographic cutoff.
A defining feature of the Type~B construction is that this cutoff surface is
chosen to lie at constant $\rho$, i.e.\ it follows the AdS$_2$ slicing of
AdS$_3$.
As a result, its induced metric is
\begin{equation}
ds^2_{\mathrm{ind}}
=
\ell^2\cosh^2\!\left(\frac{\rho_c}{\ell}\right)\frac{d\tau^2+dy^2}{y^2},
\label{eq:TypeB_induced_AdS2}
\end{equation}
which is \emph{exactly} AdS$_2$.

This geometric choice has an immediate and crucial consequence.
Because the cutoff surface itself is an AdS$_2$ spacetime, the physical BCFT
boundary coincides with the conformal boundary of AdS$_2$, located at $y=0$.
The BCFT boundary is therefore asymptotic rather than located at finite proper
distance.
In particular, the cutoff surface at $\rho=\rho_c$ intersects the EOW brane only
at $y\to0$.
There is no finite intersection point at which a boundary endpoint could move.

Thus, in Type~B the BCFT boundary is \emph{geometrically fixed} by construction:
it is identified with the asymptotic boundary of the induced AdS$_2$ geometry.
This property follows directly from the choice of cutoff along AdS$_2$ slices
and does not rely on any dynamical assumption about the boundary stress tensor.

The vanishing of the displacement operator should therefore be understood as a
\emph{consequence} of this geometric setup.
Since the boundary lies at the AdS$_2$ conformal boundary, there is no finite
normal deformation of the boundary embedding that preserves the background
geometry.
Correspondingly, the normal--normal component of the stress tensor vanishes,
\begin{equation}
T_{xx}\big|_{\Sigma}=0,
\end{equation}
and the displacement operator $\mathcal D$ is identically zero.
This reflects the absence of any boundary-localized flow or boundary motion in
the Type~B realization.

\begin{figure}
	\centering
	\includegraphics[width=12cm,height=3.7cm]{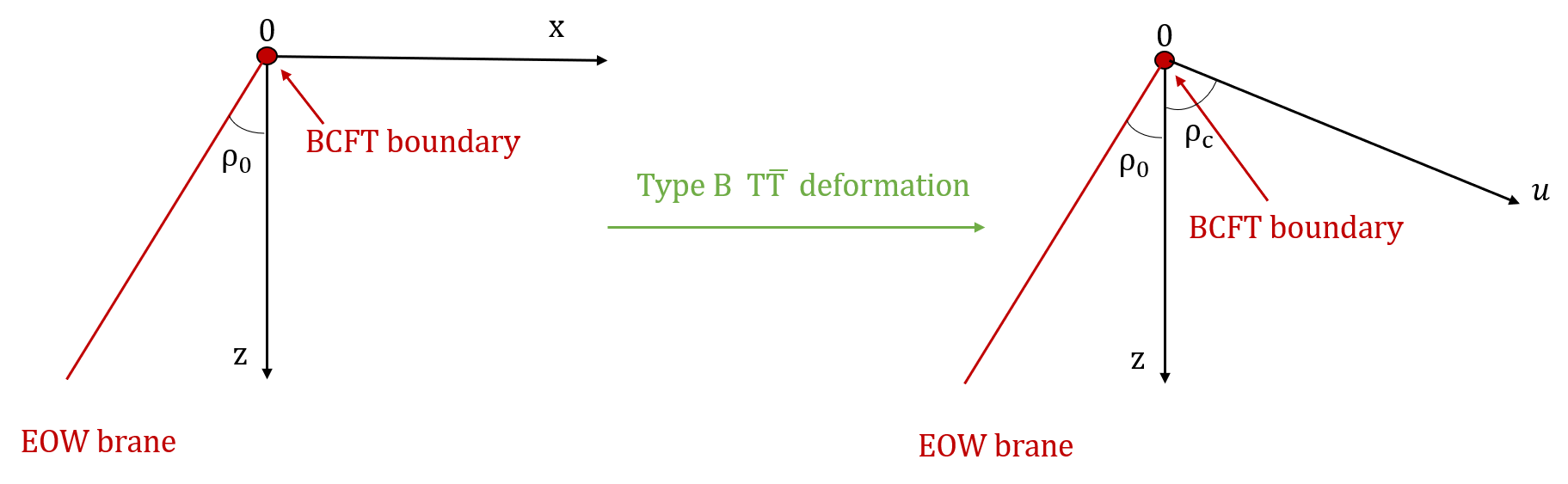}\\
	\caption{Bulk picture of the Type~B deformation in the pure AdS case.
	The bulk region is the slab $\rho_0\le\rho\le\rho_c$ in AdS$_2$ slicing.
	The Dirichlet surface at $\rho=\rho_c$ has an induced AdS$_2$ metric and
	intersects the EOW brane only at the asymptotic boundary $y\to0$, so the
	BCFT boundary remains fixed.}
	\label{B}
\end{figure}

The bulk picture of the Type~B construction is shown in Fig.~\ref{B}.

We now compute the entanglement entropy of an interval $A$ in the Type~B
realization.
Because the induced geometry on the cutoff surface is AdS$_2$ and the BCFT
boundary is fixed at the AdS$_2$ conformal boundary $y=0$, the setup is
conceptually simpler than in Type~A.
There is no boundary displacement and no effective shortening of the interval;
all effects of the deformation are encoded in the background geometry itself.

We consider an interval $A$ at fixed Euclidean time $\tau$,
\begin{equation}
A:\qquad y\in[\,0,\,y_B\,],
\label{eq:TypeB_interval}
\end{equation}
anchored at the physical boundary $y=0$ of AdS$_2$.
The coordinate $y_B$ is measured with respect to the induced metric
\eqref{eq:TypeB_induced_AdS2} on the Dirichlet surface at $\rho=\rho_c$.

\begin{figure}
	\centering
	\includegraphics[width=6cm,height=4cm]{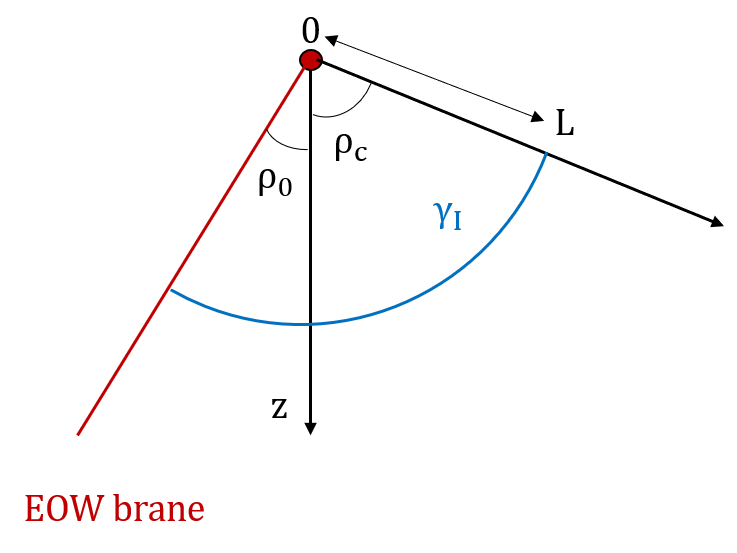}\\
	\caption{RT surface in the Type~B realization for pure AdS.
	The geodesic extends from the endpoint of the interval on the AdS$_2$
	Dirichlet boundary at $\rho=\rho_c$ to the EOW brane at $\rho=\rho_0$,
	without any boundary displacement.}
	\label{BRT}
\end{figure}

On the gravity side, the Ryu--Takayanagi surface $\gamma_A$ is a geodesic in the
bulk slab \eqref{eq:TypeB_slab_pureAdS} that starts at the endpoint
$(\rho_c,y_B)$ on the cutoff surface and ends on the EOW brane at $\rho=\rho_0$.
Because the geometry is translationally invariant along $\tau$, the geodesic
lies entirely in a constant-$\tau$ slice.
Evaluating its regulated length gives
\begin{equation}
S(A)
=
\frac{\mathrm{Area}(\gamma_A)}{4G_N}
=
\frac{\ell}{4G_N}
\log\!\left(\frac{2y_B}{\epsilon_y}\right)
+
\frac{\rho_0}{4G_N},
\label{eq:TypeB_EE_gravity}
\end{equation}
where $\epsilon_y$ is the short-distance cutoff in the AdS$_2$ coordinate $y$.
The second term is precisely the boundary entropy $S_{\mathrm{bdy}}$ given in
\eqref{eq:Sbdy_rho0}.
The corresponding RT configuration is illustrated in
Fig.~\ref{BRT}.

From the field-theory perspective, the result \eqref{eq:TypeB_EE_gravity} is
exactly what one expects for a BCFT defined on AdS$_2$.
The entanglement entropy of an interval extending from the boundary to $y=y_B$
in the vacuum state is
\begin{equation}
S_{\mathrm{BCFT}}(y_B)
=
\frac{c}{6}\log\!\left(\frac{2y_B}{\epsilon_y}\right)
+
S_{\mathrm{bdy}},
\label{eq:TypeB_EE_field}
\end{equation}
with no additional $\lambda$-dependent corrections.
Using $c=\frac{3\ell}{2G_N}$ together with \eqref{eq:Sbdy_rho0}, this expression
agrees exactly with the gravity result \eqref{eq:TypeB_EE_gravity}.

This analysis makes clear that the absence of boundary displacement in Type~B is
not a dynamical assumption but a geometric consequence of defining the cutoff
surface along AdS$_2$ slices.
The vanishing of the displacement operator $\mathcal D$ and the absence of a
boundary-localized flow are therefore consistency conditions, not independent
inputs.

\subsection{Finite temperature case: BTZ}

We again start from the Euclidean BTZ geometry
\eqref{eq:btz_metric}.
In contrast to the Type~A construction, the defining requirement of the Type~B
realization is that the physical BCFT boundary be preserved.
Geometrically, this means that the cutoff surface must share the same asymptotic
conformal boundary as the undeformed BCFT and therefore cannot be taken at
constant radial position.

Instead, we introduce a nontrivial cutoff trajectory in the bulk,
\begin{equation}
x=x(z),
\label{eq:TypeB_cutoff_ansatz}
\end{equation}
chosen such that the induced metric on the cutoff surface is intrinsically
AdS$_2$.
This geometric requirement ensures that the cutoff surface itself possesses an
asymptotic conformal boundary at infinite proper distance.
As a result, the BCFT boundary is fixed by construction and cannot be displaced
by any finite deformation.

Pulling back the BTZ metric \eqref{eq:btz_metric} onto the surface
\eqref{eq:TypeB_cutoff_ansatz}, one finds the induced line element
\begin{equation}
ds^2_{\mathrm{ind}}
=
\ell^2\left(
\frac{f(z)}{z^2}d\tau^2
+
\frac{1+f(z)x'(z)^2}{f(z)z^2}dz^2
\right),
\label{eq:TypeB_induced_metric_general}
\end{equation}
where $f(z)=1-(z/z_H)^2$.
We now impose that this two-dimensional metric have constant negative curvature.
Specifically, we require
\begin{equation}
\mathcal R_{\mathrm{ind}}
=
-\frac{2(1-\ell^2T_c^2)}{\ell^2},
\qquad
0\le T_c<\frac{1}{\ell},
\label{eq:TypeB_Ricci_target}
\end{equation}
so that the cutoff surface is locally AdS$_2$, with curvature scale controlled
by the parameter $T_c$.

Solving the resulting differential equation for the embedding function $x(z)$
yields
\begin{equation}
x(z)
=
z_H\,\mathrm{arcsinh}\!\left(
\frac{\ell T_c\,z}{z_H\sqrt{1-\ell^2T_c^2}}
\right),
\label{eq:TypeB_x_of_z}
\end{equation}
which satisfies $x(0)=0$.
Thus the cutoff surface intersects the asymptotic AdS$_3$ boundary at the same
location as the undeformed BCFT, consistently preserving the position of the
physical boundary.

\begin{figure}
	\centering
	\includegraphics[width=13cm,height=3.7cm]{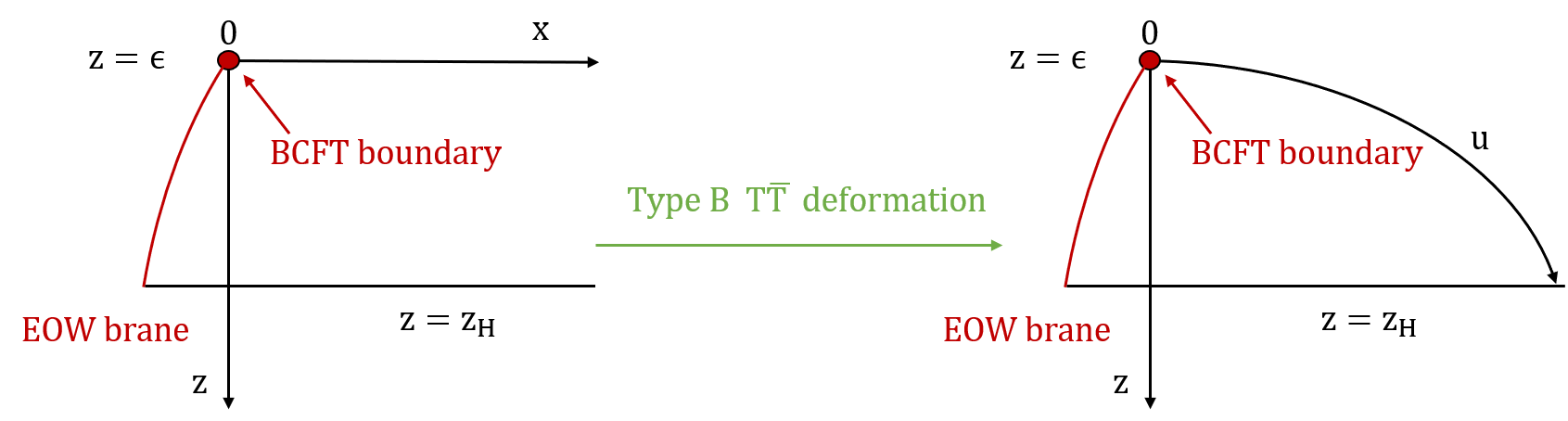}\\
	\caption{Bulk picture of the Type~B realization in the BTZ geometry.
	The cutoff surface follows an AdS$_2$ trajectory and intersects the
	asymptotic boundary at a fixed point, preserving the BCFT boundary.}
	\label{B1}
\end{figure}

Substituting \eqref{eq:TypeB_x_of_z} back into
\eqref{eq:TypeB_induced_metric_general}, the induced metric takes the explicit
form
\begin{equation}
\begin{aligned}
ds^2_{\mathrm{ind}}
&=
\ell^2\left(
\left(\frac{1}{z^2}-\frac{1}{z_H^2}\right)d\tau^2
+
\frac{\ell^2z_H^4}{z^2(z_H^2-z^2)\left(\ell^2T_c^2(z^2-z_H^2)+z_H^2\right)}dz^2
\right),
\end{aligned}
\label{eq:TypeB_induced_metric_explicit}
\end{equation}
which indeed has constant curvature \eqref{eq:TypeB_Ricci_target}.
The corresponding bulk geometry is shown in Fig.~\ref{B1}.

We now compute the entanglement entropy in the Type~B realization and demonstrate
explicit agreement between the bulk gravitational computation and the intrinsic
BCFT description on the cutoff surface.

\begin{figure}
  \centering
  \includegraphics[width=6cm,height=4cm]{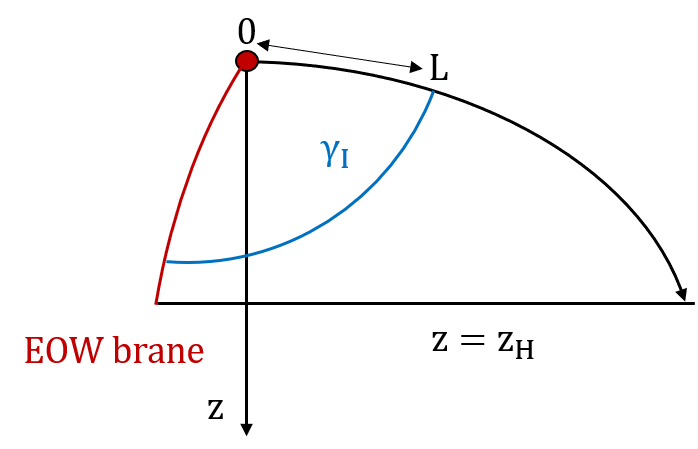}\\
  \caption{RT surface relevant for the gravity computation in the BTZ geometry
  with an AdS$_2$ cutoff trajectory.
  The extremal surface extends from the endpoint of the boundary interval on the
  AdS$_2$ Dirichlet boundary to the end-of-the-world brane, following the cutoff
  trajectory without inducing any boundary displacement.}
  \label{B1RT}
\end{figure}

We begin with the gravity computation.
The relevant Ryu--Takayanagi (RT) surface in the BTZ geometry with an AdS$_2$
cutoff trajectory is shown in Fig.~\ref{B1RT}.

The RT surface is a bulk geodesic that connects the endpoint of the interval on
the cutoff surface to the EOW brane.
Because the geometry is static, the geodesic lies in a constant Euclidean time
slice and may be evaluated using standard embedding-coordinate techniques in
AdS$_3$.

The endpoint of the interval on the cutoff surface is located at
\begin{equation}
(\tau,x,z)
=
\left(
0,\;
z_H\,\operatorname{arcsinh}\!\left(
\frac{\ell T_c\, z_L}{z_H\sqrt{1-\ell^2 T_c^2}}
\right),\;
z_L
\right),
\end{equation}
while the second endpoint lies on the EOW brane and may be parametrized as
\begin{equation}
(\tau,x,z)
=
\left(
0,\;
-\,z_H\,\operatorname{arcsinh}\!\left(
\frac{\ell T\, z'}{z_H\sqrt{1-\ell^2 T^2}}
\right),\;
z'
\right).
\end{equation}
The regulated geodesic length $d(z')$ depends on the intersection point $z'$
with the brane.
Extremizing $d(z')$ with respect to $z'$ yields
\begin{equation}
z'_{\mathrm{min}}
=
z_L z_H
\sqrt{
\frac{1-\ell^2 T_c^2}
{z_H^2-\ell^2 T_c^2+\ell^2 T^2\left(z_L^2-z_H^2\right)}
}.
\end{equation}
Substituting $z'_{\mathrm{min}}$ back into the geodesic length and dividing by
$4G_N$, the holographic entanglement entropy takes the form
\begin{equation}
S(B)
=
\frac{\ell}{4G_N}\,
\operatorname{arccosh}\!\left(\frac{1}{\sqrt{1-\ell^2 T^2}}\right)
+
\frac{\ell}{4G_N}\,
\operatorname{arccosh}\!\left(\frac{1}{\sqrt{1-\ell^2 T_c^2}}\right).
\end{equation}
The first term depends only on the EOW brane tension and reproduces the boundary
entropy $S_{\mathrm{bdy}}$, while the second term arises entirely from the cutoff
surface and encodes the intrinsic curvature scale of the AdS$_2$ geometry on
which the deformed BCFT is defined.

We now turn to the field-theory computation.
In the Type~B realization, the BCFT lives on the cutoff surface itself, whose
induced metric was given in \eqref{eq:TypeB_induced_metric_explicit}.
To make contact with standard BCFT formulas, it is convenient to introduce a
coordinate $y$ adapted to the AdS$_2$ slicing of the cutoff surface,
\begin{equation}
y
=
z_H\,
\operatorname{arctanh}\!
\left(
\frac{z}{\sqrt{z_H^2+\ell^2 T_c^2\left(z^2-z_H^2\right)}}
\right).
\end{equation}
In terms of $y$, the induced metric becomes manifestly conformally flat,
\begin{equation}
ds^2_{\mathrm{ind}}
=
\frac{\ell^2}{z_H^2\sinh^2\!\left(\frac{y}{z_H}\right)}
\frac{-d\tau^2+dy^2}{1-\ell^2 T_c^2}
\equiv
\Omega^{-2}(y)\,ds^2_{\mathrm{flat}},
\end{equation}
where
\begin{equation}
ds^2_{\mathrm{flat}}
=
\frac{-d\tau^2+dy^2}{1-\ell^2 T_c^2},
\end{equation}
and $\Omega(y)$ denotes the Weyl factor.
The appearance of the overall factor $(1-\ell^2 T_c^2)^{-1}$ reflects the fact
that the intrinsic AdS$_2$ curvature scale sets an effective UV cutoff along the
$y$ direction.

For a thermal BCFT defined on the flat background $ds^2_{\mathrm{flat}}$, the
entanglement entropy of an interval $y\in[0,y_L]$ is
\begin{equation}
S_{\mathrm{flat}}
=
\frac{c}{6}
\log\!\left(
\frac{2z_H}{\sqrt{1-\ell^2 T_c^2}}
\sinh\!\frac{y_L}{z_H}
\right)
+
S_{\mathrm{bdy}} .
\end{equation}
Under a Weyl rescaling $ds^2\to\Omega^{-2}(y)ds^2$, the short-distance cutoff
transforms as $\epsilon_y\to\Omega(y)\epsilon_y$.
Taking this into account, the entanglement entropy on the physical induced
geometry becomes
\begin{equation}
\begin{aligned}
S(B)
&=
\frac{c}{6}
\log\!\left(
\frac{2z_H}{\Omega(y_L)\sqrt{1-\ell^2 T_c^2}}
\sinh\!\frac{y_L}{z_H}
\right)
+
S_{\mathrm{bdy}}
\\
&=
\frac{c}{6}
\operatorname{arccosh}\!\left(\frac{1}{\sqrt{1-\ell^2 T_c^2}}\right)
+
\frac{c}{6}
\operatorname{arccosh}\!\left(\frac{1}{\sqrt{1-\ell^2 T^2}}\right),
\end{aligned}
\end{equation}
where in the final step we used the explicit form of the Weyl factor and the
relation between the boundary entropy and the EOW brane tension.
Using $c=\frac{3\ell}{2G_N}$, this result matches the holographic expression
exactly.

This agreement provides a nontrivial consistency check of the Type~B
construction.
Unlike in Type~A, the $T\bar T$ deformation in Type~B does not manifest itself as a
boundary displacement.
Instead, all effects of the deformation are encoded in the intrinsic AdS$_2$
geometry of the cutoff surface, while the physical BCFT boundary remains fixed at
the AdS$_2$ conformal boundary.

\section{Conclusion and discussion}
\label{con}

In this work we have analyzed the $T\bar T$ deformation of boundary conformal
field theories from an intrinsic field-theoretic perspective and through its
holographic realizations in $\mathrm{AdS}_3/\mathrm{BCFT}_2$.
Our starting point was the formulation of the $T\bar T$ deformation as a mixed
asymptotic boundary condition in $\mathrm{AdS}_3$, which yields an exact
quadratic trace relation for the stress tensor without introducing a finite
radial cutoff.
This intrinsic formulation provides a nonperturbative and
representation-independent framework for the deformation and serves as the
conceptual backbone of the present analysis.

When restricted to a BCFT without independent boundary degrees of freedom, the
intrinsic $T\bar T$ deformation acquires a genuinely boundary-localized
character.
Imposing reflective boundary conditions collapses the bulk composite operator
to a purely boundary contribution governed by the displacement operator.
We derived an exact boundary flow equation and integrated it in closed form,
obtaining a universal one-dimensional irrelevant boundary action.
This construction makes explicit that the $T\bar T$ deformation does not
introduce new boundary degrees of freedom.
Instead, it reorganizes existing boundary data into a nonlinear functional of
the displacement operator, fully determined by BCFT Ward identities.
From this perspective, the deformation modifies the kinematic response of the
boundary without enlarging its operator content.

A central conceptual result of our field-theoretic analysis is the precise
equivalence between two descriptions of the boundary deformation.
In the fixed-boundary representation, the deformation is encoded through an
induced boundary action added to the generating functional.
In the moving-boundary representation, the same deformation is realized as a
$\lambda$-dependent shift of the boundary embedding, with the displacement
operator acting as the conjugate variable.
We demonstrated that these two descriptions are related by a variational
(Legendre-type) transformation and therefore constitute two parameterizations
of the same intrinsic boundary-localized flow.
This equivalence concerns the variational problem and the resulting space of
classical solutions, rather than the physical motion of an independent boundary
degree of freedom, and should be clearly distinguished from differences that
arise from particular holographic realizations.

\paragraph{Intrinsic versus cutoff-based descriptions.}

An important conceptual aspect of the present framework is the distinction
between intrinsic and cutoff-based descriptions of the $T\bar T$ deformation.
In the intrinsic formulation adopted here, the deformation is defined as a
modification of the asymptotic variational principle and does not introduce new
boundary degrees of freedom.
The displacement operator that governs the boundary-localized flow is therefore
an intrinsic operator of the undeformed BCFT.
Its role is to mediate geometric reparametrizations of the boundary data rather
than to undergo renormalization or operator mixing along the $\lambda$ flow.
All dependence on the deformation parameter is encoded in expectation values
and in the geometric relations implied by the variational principle.

This viewpoint differs from approaches in which the deformation is implemented
by introducing a finite radial cutoff and treating the cutoff surface itself as
a dynamical object.
In such descriptions, geometric motion may appear explicitly and can obscure
which aspects of the deformation are intrinsic and which depend on a particular
choice of holographic realization.
By contrast, the intrinsic perspective isolates the deformation at the level of
boundary Ward identities and makes transparent the sense in which different bulk
geometries encode the same underlying boundary-localized physics.

On the holographic side, we identified two inequivalent bulk realizations of
the same intrinsic deformation, referred to as Type~A and Type~B.
Both realizations implement the identical mixed asymptotic boundary condition
that defines the $T\bar T$ deformation, but they differ in how boundary data are
encoded geometrically in the bulk.

In the Type~A realization, the Dirichlet cutoff surface is taken to be rigid,
while the end-of-the-world brane continues to satisfy its Neumann boundary
condition.
Varying the deformation parameter causes the intersection point between the
cutoff surface and the brane to move along the boundary direction.
When described in terms of the induced geometry on the cutoff surface, this
motion appears as a finite displacement of the physical BCFT boundary.
In this case the displacement operator is nonvanishing, and the boundary motion
provides a direct geometric representation of the intrinsic
boundary-localized flow.

In the Type~B realization, by contrast, the cutoff surface is chosen such that
its induced metric is asymptotically $\mathrm{AdS}_2$.
With this choice, the cutoff surface intersects the end-of-the-world brane only
at asymptotic infinity.
The physical BCFT boundary therefore coincides with the conformal boundary of
an $\mathrm{AdS}_2$ geometry and is geometrically pinned, in the sense that no
finite boundary location exists whose embedding could be shifted by varying the
deformation parameter.
As a result, the normal--normal component of the stress tensor vanishes at the
boundary, the displacement operator is identically zero, and no explicit
boundary motion is generated.
Importantly, this absence of boundary displacement should be understood as a
consequence of the asymptotic $\mathrm{AdS}_2$ structure of the cutoff geometry
rather than as an independent dynamical assumption.
In this realization the intrinsic boundary-localized flow is not absent, but
rather is geometrically frozen and absorbed entirely into the background
geometry of the cutoff surface.

This comparison clarifies an important conceptual point.
Type~A and Type~B do not correspond to different deformations of the BCFT, nor do
they map directly onto the fixed- versus moving-boundary representations in
field theory.
In particular, Type~B cannot be obtained from Type~A by a coordinate
redefinition or boundary reparametrization, since the existence of a finite
intersection point between the cutoff surface and the end-of-the-world brane is
a geometric invariant.
Instead, the two constructions are genuinely distinct holographic realizations
of the same intrinsic boundary-localized deformation, distinguished only by how
boundary data are encoded in bulk geometric variables.

We used entanglement entropy, both at zero and finite temperature, as a
quantitative probe of these two holographic realizations.
The agreement with BCFT expectations provides a nontrivial consistency check
of the intrinsic boundary-localized flow and demonstrates that universal
observables are insensitive to the specific geometric realization chosen in
the bulk.
More broadly, our results highlight the usefulness of entanglement entropy as a
diagnostic tool for disentangling intrinsic properties of irrelevant
deformations from features that depend on particular holographic
implementations.

There are several natural directions for future work.
It would be interesting to extend the intrinsic analysis developed here to
other irrelevant deformations and to investigate whether analogous
boundary-localized structures arise more generally, in particular in theories
with independent boundary degrees of freedom.
Another direction is the study of higher-dimensional generalizations, where
boundary Ward identities and displacement operators possess a richer
structure.
Finally, a more systematic classification of admissible holographic cutoff
geometries and their relation to intrinsic mixed boundary conditions may
provide further insight into how different bulk realizations encode the same
ultraviolet deformation.

We hope that the intrinsic perspective adopted in this work helps clarify the
role of boundaries in $T\bar T$-deformed theories and provides a useful
framework for future investigations of irrelevant deformations in holographic
and boundary quantum field theories.
\section*{Acknowledgments}
\section*{Acknowledgements}

Feiyu Deng thanks Make Yuan and Zhi Wang for useful discussions.
He is particularly grateful to Hao Ouyang for pointing out an inconsistency in
the original dimensional analysis, which led to a clarification and correction
of the boundary-localized formulation.
This work is supported by the National Natural Science Foundation of China (NSFC)
under Project No.~12547135.
\bibliographystyle{JHEP}
\bibliography{qes=hee}
\appendix

\end{document}